\documentclass[paper,notoc]{JHEP3}
\usepackage{epsfig,cite}
\usepackage{amsbsy} 
\usepackage{graphicx}

\def\beq{\begin{equation}}   
\def\eeq{\end{equation}}
\def\bea{\begin{eqnarray}}  
\def\eea{\end{eqnarray}}

\title{The fully differential decay rate of a Higgs boson to bottom-quarks at NNLO in QCD}

\author{Charalampos Anastasiou\\
  Institute for Theoretical Physics, ETH Zurich,
  8093 Zurich, Switzerland\\
  E-mail: \email{babis@phys.ethz.ch}}

\author{Franz Herzog\\
  Institute for Theoretical Physics, ETH Zurich,
  8093 Zurich, Switzerland\\
  E-mail: \email{fherzog@itp.phys.ethz.ch}}

\author{Achilleas Lazopoulos\\
  Institute for Theoretical Physics, ETH Zurich,
  8093 Zurich, Switzerland\\
  E-mail: \email{lazopoli@itp.phys.ethz.ch}}

\abstract{
The decay of  a light Higgs  boson  to bottom quarks is dominant and can be exploited 
for the discovery of the Higgs particle and the measurement of its properties at the LHC  and future collider experiments. 
We perform a first computation of  the fully differential decay at next-next-to-leading order  in perturbative QCD.  
We employ a novel method  of non-linear mappings  for the treatment of  singularities in the radiative processes which 
contribute  to the decay width. This constitutes the first physical application of  the  method. }

\keywords{QCD, NLO, NNLO, LHC, Tevatron}

\preprint{}

\begin{document}

\section{Introduction}
\label{sec:introduction}

Direct searches and electroweak precision tests suggest that   
a  Standard Model Higgs boson is light, and that it should decay predominantly into a bottom quark 
($b\bar{b}$) pair. Inclusive searches at the LHC of a Higgs resonance in the bottom-pair invariant mass 
distribution are unfortunately hampered  by a very large QCD background. Direct searches are focusing on other 
decay channels,  such as $H \to \gamma \gamma$,  with a much smaller branching ratio.  In view of this, 
perturbative corrections to the $H \to b \bar{b}$ decay  are most interesting for the inclusive rate, due to its 
contribution  to the total decay width and the branching ratios of other decays.
A remarkable effort  has been made for a precise calculation of  $\Gamma_{H \to b \bar{b}}$ 
and now  QCD corrections  are known to $\mathcal{O}(\alpha_s^4)$ \cite{Baikov:2005rw,Gorishnii:1983cu,Gorishnii:1990zu,Gorishnii:1991zr,Becchi:1980vz, Sakai:1980fa, Inami:1980qp, Chetyrkin:1996sr,Chetyrkin:1997vj}. 

Recently, confidence has grown that the $H \to b \bar{b}$ decay can be observed at the LHC directly 
in events  where the Higgs boson is  produced in  association with a  massive vector boson $(W,Z)$~\cite{Butterworth:2008iy}  or a 
$t\bar t$ pair~\cite{Plehn:2009rk}.
Backgrounds from $t\bar t$,$V+jj$ and multi-jet production are challenging. However, the  excellent $b-$jet tagging  of  
the ATLAS and CMS detectors as well as sophisticated selections of jet events described in~\cite{Butterworth:2008iy,Plehn:2009rk}, 
render these channels hopeful, as explicitly demonstrated by the ATLAS collaboration in a full detector simulation analysis for the $ZH$ case~\cite{ATLAS:ZH}. For this channel, further suppression of the background can be obtained with the sub-jet algorithms of~\cite{Soper:2010xk}.

These search strategies rely on a selection of phase space corners which are rich in potential Higgs events.  
An accurate modeling of QCD radiation is necessary in order to assess the efficiency of these selections.  
This motivates the computation of the fully differential $H\rightarrow b\bar b$ decay rate at 
next-to-next-to-leading-order (NNLO)  in perturbative QCD.

Next-to-leading-order (NLO) computations~\cite{Bredenstein:2010rs, Denner:2010jp, Melia:2010bm, KeithEllis:2009bu, Berger:2010zx,  Berger:2010vm, Berger:2009ep, Ita:2011wn, Bevilacqua:2009zn, Bevilacqua:2010ve, Bevilacqua:2010qb,Greiner:2010ci,Frederix:2011qg} are currently performed with very well automated methods~\cite{ Berger:2008sj, Ellis:2008ir, vanHameren:2009dr, Hirschi:2011pa, Frederix:2008hu,Frederix:2009yq, Gleisberg:2007md, Binoth:2008uq, Mastrolia:2010nb}. 
Obtaining fully differential cross-sections and decay rates at one order  higher in the perturbative  expansion 
requires the solution of new challenging problems.  In the last decade rapid progress~\cite{Kosower:1997zr,GehrmannDeRidder:2003bm,GehrmannDeRidder:2005cm, Daleo:2006xa, GehrmannDeRidder:2009fz, Daleo:2009yj, Glover:2010im, Boughezal:2010mc, Abelof:2011jv, Gehrmann:2011wi, Weinzierl:2003fx, Frixione:2004is, Somogyi:2005xz, Somogyi:2006da, Somogyi:2006db, Bolzoni:2009ye, Bolzoni:2010bt, Somogyi:2008fc,Aglietti:2008fe,Czakon:2010td,Czakon:2011ve,Anastasiou:2003gr,Catani:2007vq, Binoth:2000ps,Binoth:2004jv} with important results~\cite{Ferrera:2011bk, Catani:2009sm, Melnikov:2006kv, Anastasiou:2003yy, Anastasiou:2003ds, Anastasiou:2002qz, Anastasiou:2004xq, Anastasiou:2005qj, Grazzini:2008tf, Anastasiou:2004qd, GehrmannDeRidder:2004tv, GehrmannDeRidder:2007hr, GehrmannDeRidder:2007jk} has been achieved.   
However, further developments of methods and new ideas are necessary for efficient cancelations of infrared singularities and  
evaluations of novel two-loop amplitudes in more complicated LHC processes. 

In Ref.~\cite{Anastasiou:2010pw} we introduced a new  method to extract the singularities of multi-dimensional loop 
and phase space integrals which emerge at NNLO. The  method employs non-linear transformations for the 
factorization of overlapping singularities which in turn  allow  for an efficient numerical evaluation. 
In this article, we  demonstrate the method in a realistic physical example. We compute  for the first time the  
fully  differential decay rate for $H \to b \bar{b}$ at NNLO.  

The  paper is organized  as  follows.  In Section~\ref{sec:notation} we fix our notation and set up the calculation. 
In Section~\ref{sec:matrix elements} we present the matrix elements for the processes which contribute to the decay width 
through NNLO. In Section~\ref{sec:phase space} we describe the parameterization of the phase space for the final state partons.  
In Section~\ref{sec:real-virtual} and  Section~\ref{sec:double-real} we explain how our method leads to a simple factorizable form 
for the singularities in the integrals which emerge at NNLO. In Section~\ref{sec:numerics} we present the inclusive width and 
examples  of differential observables with our numerical code. 


\section{Notation}
\label{sec:notation}

We  compute  the partial width $\Gamma_{H \to b\bar{b}X}[{\cal J}]$ through NNLO in perturbative QCD 
for any infrared safe observable ${\cal J}$, such as an appropriate jet-algorithm,  
in the decay of a Higgs  boson to a pair of bottom-quarks, 
\begin{equation}
\label{eq:process} 
H \to  b + \bar{b} + X. 
\end{equation}
 We  consider that the Higgs boson couples to a bottom-quark pair with an unrenormalized 
Yukawa coupling $y_b^B$. Due to the largely different  mass  of the Higgs boson, $m_H$, and the  small bottom-quark, $m_b$,  
we treat the latter as independent of the Yukawa coupling and set it to zero, $m_b =0$, in matrix elements and phase space integrals.  
In addition, we  neglect top-quarks both in virtual corrections  and in the renormalization procedure and consider $n_f =5$ light quark flavors. 
For the renormalization of ultraviolet (UV) divergences we work in the $\overline{\rm{MS}}$-scheme where the bare couplings, $y_b^B$, 
and $g_s^B=\sqrt{4\pi\alpha_s^B}$, are related to their renormalized counterparts  
through the following relations:
\begin{eqnarray}
y_b^B & = & y_b \Bigg[ 1-\frac{\alpha_s}{4\pi}\frac{3C_F}{\epsilon}  +\left(\frac{\alpha_s}{4\pi}\right)^2 \Big[ -\frac{1}{\epsilon^2}\left( 2T_Fn_f-\frac{9}{2}C_F-\frac{11}{2}C_A \right) \nonumber  \\
& &  +\frac{1}{\epsilon}\left( \frac{5T_F}{3}n_f-\frac{3C_F}{4} - \frac{97}{12}C_A \right)  \Big] C_F + \mathcal{O}(\alpha_s^3) \Bigg]  \\
\alpha_s^B & = & \alpha_s  \left( \frac{\mu^2 e^{\gamma_E}}{4\pi} \right)^{\epsilon} \left[ 1-\frac{\alpha_s}{4\pi}\frac{1}{\epsilon}
\left[ \frac{11C_A}{3}  -\frac{4T_Fn_f}{3} \right]+\mathcal{O}(\alpha_s^2)   \right] . 
\end{eqnarray}
The renormalized strong coupling and Higgs Yukawa coupling depend  implicitly  on the renormalization scale $\mu$, 
\beq 
\alpha_s \equiv \alpha_s(\mu)  \quad y_b \equiv y_b(\mu). 
\eeq
The SU(N) Casimirs are given by 
\beq
C_A=N, \quad C_F=\frac{N^2-1}{2N},  \quad T_F=\frac{1}{2}, 
\eeq
where $N=3$ is the number of quark colors.  

Through NNLO, the  partial decay width receives contributions from the following  partonic processes:  
\begin{itemize}
\item  $H(p_H) \to b(p_1) \bar{b}(p_2)$  at two loops, 
\item  $H(p_H) \to b(p_1) \bar{b}(p_2) g(p_3) $  at one loop and
\item $H(p_H) \to b(p_1) \bar{b}(p_2)g(p_3)g(p_4), 
b(p_1) \bar{b}(p_2) b(p_3) \bar{b}(p_4), 
b(p_1) \bar{b}(p_2) q(p_3) \bar{q}(p_4) $   
with  $q \neq b$ at tree level,    
\end{itemize} 
where in parenthesis we indicate the four-momenta of the corresponding particles. 
The  width is  the  sum 
\begin{eqnarray}
\Gamma_{H \to b\bar{b}X}[{\cal J}] &=& \Gamma_{H \to b\bar{b}}[{\cal J}]  \nonumber \\ 
                           &+& \Gamma_{H \to b\bar{b}g}[{\cal J}] \nonumber \\ 
                            &+& \Gamma_{H \to b\bar{b}gg}[{\cal J}] + (n_f-1) \cdot \Gamma_{H \to b\bar{b}qq}[{\cal J}] 
                            +\Gamma_{H \to b\bar{b}b \bar{b}}[{\cal J}] \nonumber \\
                            &+& {\cal O}(\alpha_s^3), 
\end{eqnarray}
where 
\begin{eqnarray}
\Gamma_{H \to b\bar{b}}[{\cal J}] &=& 
\frac{1}{2m_H} \int d\Phi_2 \sum_{{\rm spin}} \left| {\cal M}_{H \to b(p_1) + \bar{b}(p_2)} \right|^2  
{\cal J}(p_1, p_2), \nonumber \\
\Gamma_{H \to b\bar{b} g}[{\cal J}] &=& 
\frac{1}{2m_H} \int d\Phi_3 \sum_{{\rm spin}} \left| {\cal M}_{H \to b(p_1) + \bar{b}(p_2)+g(p_3)} \right|^2  
{\cal J}(p_1, p_2, p_3), \nonumber \\
\Gamma_{H \to b\bar{b} g g}[{\cal J}] &=& 
\frac{1}{2m_H} \int d\Phi_4 \sum_{{\rm spin}} \left| {\cal M}_{H \to b(p_1) + \bar{b}(p_2)+g(p_3)+g(p_4)} \right|^2  
\cdot \frac{1}{2!} \cdot {\cal J}(p_1, p_2, p_3, p_4), \nonumber\\
\Gamma_{H \to b\bar{b} b  \bar{b}}[{\cal J}] &=& 
\frac{1}{2m_H} \int d\Phi_4 \sum_{{\rm spin}} \left| {\cal M}_{H \to b(p_1) + \bar{b}(p_2)+b(p_3)+\bar{b}(p_4)} \right|^2  
\cdot \frac{1}{(2!)^2} \cdot {\cal J}(p_1, p_2, p_3, p_4),	 \nonumber\\
\Gamma_{H \to b\bar{b} q  \bar{q}}[{\cal J}] &=& 
\frac{1}{2m_H} \int d\Phi_4 \sum_{{\rm spin}} \left| {\cal M}_{H \to b(p_1) + \bar{b}(p_2)+q(p_3)+\bar{q}(p_4)} \right|^2  
 {\cal J}(p_1, p_2, p_3, p_4).  
\end{eqnarray}
In the above, we integrate the appropriate matrix elements computed in conventional dimensional regularization (CDR) 
over the phase space  in $D=4-2\epsilon$ dimensions of  the final-state partons and  the function ${\cal J}$ 
defining an infrared safe observable. The phase space measure reads
\begin{equation}
d \Phi_n \equiv (2 \pi)^D \delta^{(D)} 
\left(p_H - \sum_{i=1}^n p_i \right) \prod_{i=1}^n \frac{d^Dp_i}{\left(2 \pi\right)^{D-1}} \delta^{(+)}(p_i^2).
\end{equation} 
Lorentz invariants are defined as 
\begin{equation}
m_H^2 = p_H^2,  \quad 
s_{ij} = 2p_i.p_j,  \quad s_{ijk} = s_{ij}+s_{ik}+s_{jk}. 
\end{equation}
Regarding the analytic continuations of complex functions presented in this paper the positive
``epsilon prescription'' must be used, since all occurring mass scales are external,i.e.
\begin{equation}
 m_H^2 \rightarrow m_H^2+i\varepsilon, \quad s_{ij} \rightarrow s_{ij}+i\varepsilon.
\end{equation}
where $\varepsilon$ is a small positive parameter.

\section{Matrix elements}
\label{sec:matrix elements}

In this section we present the square of the scattering amplitudes needed for the computation of the $H\rightarrow b\bar b$ decay width
through $\mathcal{O}(\alpha_s(\mu)^2)$. These can be easily derived  with modern methods for  computing Feynman diagrams 
and have been ingredients  of many prior calculations. For example, the same matrix elements enter the calculation of  the NNLO 
total cross-section $b \bar{b} \to H$~\cite{Harlander:2003ai}.  We have made an independent computation and present them here for completeness. 
For the generation of the matrix elements we used QGRAF \cite{Nogueira:1991ex}. For further symbolic manipulations, 
such as color and Dirac algebra,  we used FORM \cite{Vermaseren:2000nd} and MAPLE~\cite{maple}. 
We reduced one and two-loop amplitudes  to master integrals using the method of integration by parts~\cite{Chetyrkin:1981qh,Tkachov:1981wb} and 
the Laporta algorithm~\cite{Laporta:2001dd} with AIR \cite{Anastasiou:2004vj}. 

\subsection{Decay to two partons}
For the process 
$$
H(p_H)\rightarrow b(p_1)+\bar b(p_2)
$$ 
we require up to 2-loop corrections:
\beq
\left|\mathcal{M}_{H\rightarrow b\bar b} \right|^2= 2y_b(\mu)^2 m_H^2N \left[1 +\frac{\alpha_s(\mu)}{\pi}\mathcal{A}_{H\rightarrow b\bar b}^{(1)}
+\left(\frac{\alpha_s(\mu)}{\pi}\right)^2 \mathcal{A}_{H\rightarrow b\bar b}^{(2)}+\mathcal{O}(\alpha_s(\mu)^3) \right],
\eeq
where (let $\Re(z)$ shall denote the real part of $z$)
\beq
\mathcal{A}_{H\rightarrow b\bar b}^{(1)}  =  -C_F\Bigg[  \Re\left(\left( \frac{\mu^2}{-m_H^2} \right)^{\epsilon}\right) f_\epsilon+\frac{3}{2\epsilon}\Bigg]\nonumber
\eeq
with
\beq
f_\epsilon= (1-\epsilon)^2 \frac{\Gamma(1+\epsilon)\Gamma(-\epsilon)^2}{\Gamma(2-2\epsilon)}e^{\gamma \epsilon}.
\eeq
The $\mathcal{O}(\alpha_s(\mu)^2)$ contribution can be split into two pieces 
\begin{eqnarray}
\mathcal{A}_{H\rightarrow b\bar b}^{(2)}= \mathcal{A}_{H\rightarrow b\bar b}^{VV}+\mathcal{A}_{H\rightarrow b\bar b}^{V^2}
\end{eqnarray}
corresponding to the one-loop squared amplitude 
\beq
\mathcal{A}_{H\rightarrow b\bar b}^{V^2}=\frac{C_F^2}{4} \left[ \left|\frac{\mu^2}{-m_H^2}\right|^{2\epsilon} f_\epsilon^2 
			     +\frac{3}{\epsilon}\Re\left(\left( \frac{\mu^2}{-m_H^2} \right)^{\epsilon}\right) f_\epsilon  + \frac{9}{4\epsilon^2}\right].
\eeq
and the two-loop amplitude interfered with the tree amplitude
\beq
\mathcal{A}_{H\rightarrow b\bar b}^{VV}=\left[ \frac{\mathcal A^{VV}_{-4}}{\epsilon^{4}} +\frac{\mathcal A^{VV}_{-3}}{\epsilon^{3}}+\frac{\mathcal A^{VV}_{2}}{\epsilon^{2}}+\frac{\mathcal A^{VV}_{-1}}{\epsilon}+\mathcal A^{VV}_0 \right]
\eeq

where
\begin{eqnarray}
\mathcal A_0^{VV} &=& \Bigg[ 1/6\,{l_{{H}}}^{4}+{\frac {13}{72}}\,{l_{{H}}}^{3}+ \left( -6
\,\zeta \left( 2 \right) -{\frac {31}{36}} \right) {l_{{H}}}^{2}+
 \left( -{\frac {55}{108}}-{\frac {61}{24}}\,\zeta \left( 2 \right) -7
/6\,\zeta  \left( 3 \right)  \right) l_{{H}}\nonumber\\
& & 
+{\frac {455}{162}}+{
\frac {377}{72}}\,\zeta \left( 2 \right) +{\frac {263}{16}}\,\zeta
 \left( 4 \right) -{\frac {47}{36}}\,\zeta  \left( 3 \right)  \Bigg]
{C_{{F}}}^{2}\nonumber\\
& & 
+ \Bigg[ -1/36\,n_{{f}}{l_{{H}}}^{3}+{\frac {5}{36}}\,n_{{f}}{l_{{H}}}^{2}
+1/27\, \left( 9\,\zeta \left( 2 \right) -7 \right) n_{{f}}l_{{H}}\nonumber\\
& & 
-{\frac {1}{648}}\, \left( 495\,\zeta \left( 2 \right) 
-18\,\zeta  \left( 3 \right) -200 \right) n_{{f}} \Bigg] C_{{F}}\nonumber\\
& & 
+ \Bigg[ {\frac {11}{72}}\,{l_{{H}}}^{3}+ \left( 1/4\,\zeta \left( 2
 \right) -{\frac {67}{72}} \right) {l_{{H}}}^{2}
+ \left( -{\frac {13}{4}}\,\zeta  \left( 3 \right) -{\frac {11}{6}}\,\zeta \left( 2 \right) 
+{\frac {121}{108}} \right) l_{{H}}\nonumber\\
& & 
+{\frac {701}{144}}\,\zeta \left( 2
 \right) -\zeta \left( 4 \right) -{\frac {467}{648}}+{\frac {151}{72}}
\,\zeta  \left( 3 \right)  \Bigg] C_{{F}}{N}^{-1}\nonumber\\
\mathcal A_{-1}^{VV} &=& \left( -1/3\,{l_{{H}}}^{3}+3/8\,{l_{{H}}}^{2}+ \left( 6\,\zeta
 \left( 2 \right) +{\frac {31}{36}} \right) l_{{H}}-{\frac {13}{4}}\,
\zeta \left( 2 \right) +{\frac {7}{12}}\,\zeta  \left( 3 \right) -{
\frac {491}{864}} \right) {C_{{F}}}^{2}\nonumber\\
& & 
+ \left( -{\frac {5}{36}}\,n_{{
f}}l_{{H}}+{\frac {1}{432}}\, \left( 54\,\zeta \left( 2 \right) +65
 \right) n_{{f}} \right) C_{{F}}\nonumber\\
& & 
+ \left(  \left( -1/4\,\zeta \left( 2
 \right) +{\frac {67}{72}} \right) l_{{H}}-{\frac {11}{16}}\,\zeta
 \left( 2 \right) +{\frac {13}{8}}\,\zeta  \left( 3 \right) -{\frac {
961}{864}} \right) C_{{F}}{N}^{-1}\nonumber\\
\mathcal A_{-2}^{VV} &=&
\left( -3\,\zeta \left( 2 \right) -5/3\,l_{{H}}+1/2\,{l_{{H}}}^{2}+{
\frac {217}{144}} \right) {C_{{F}}}^{2}\nonumber\\
& & 
+ \left( 1/12\,n_{{f}}l_{{H}}-1
/18\,n_{{f}} \right) C_{{F}}+ \left( 1/8\,\zeta \left( 2 \right) -{
\frac {11}{24}}\,l_{{H}}+2/9 \right) C_{{F}}{N}^{-1}\nonumber\\
\mathcal A_{-3}^{VV} &=& \left( {\frac {17}{8}}-1/2\,l_{{H}} \right) {C_{{F}}}^{2}-1/8\,C_{{F}
}n_{{f}}+{\frac {11}{16}}\,{\frac {C_{{F}}}{N}}\nonumber\\
\mathcal A_{-4}^{VV} &=& 1/4\,{C_{{F}}}^{2}
\end{eqnarray}

with $l_H=\ln\left(\frac{m_H^2}{\mu^2}\right)$.

\subsection{Decay to three partons}
For the process
$$
H(p_H)\rightarrow b(p_1)+\bar b(p_2)+g(p_3)
$$ 
we require up to 1-loop corrections:
\beq
\left|\mathcal{M}_{H\rightarrow b\bar bg} \right|^2= y_b(\mu)^2 \left[ \frac{\alpha_s(\mu)}{\pi}\mathcal{A}_{H\rightarrow b\bar bg}^{(0)}
+\left(\frac{\alpha_s(\mu)}{\pi}\right)^2 \mathcal{A}_{H\rightarrow b\bar bg}^{(1)}+\mathcal{O}(\alpha_s(\mu)^3) \right].
\eeq
The tree contribution can be expressed as
\beq
\mathcal{A}_{H\rightarrow b\bar bg}^{(0)}= m_H^2N(4\pi)^2 \left( \frac{\mu^2 e^{\gamma_E}}{4\pi} \right)^{\epsilon}  \hat{P}_{qq}\left(\frac{s_{12}}{m_H^2}\right) \left[ \frac{1}{s_{13}}+\frac{1}{s_{23}} \right] 
\eeq
where 
\begin{equation}
\hat{P}_{qq}(z)=C_F\left[\frac{1+z^2}{1-z}-\epsilon(1-z)\right]
\end{equation}
is the quark gluon splitting kernel.
The one-loop correction can be expressed as

\begin{eqnarray}
&& \mathcal{A}_{H\rightarrow b\bar bg}^{(1)} = \frac{\pi^2 \left( \mu^2 e^{\gamma_E} \right)^{2\epsilon}}{(4\pi)^{\epsilon}} \times
\Bigg\{  \Bigg( 8\,{\frac { \left( s_{23}+s_{13} \right) ^{2}{\epsilon}^{2}}{s_{23}s_{13}}}\nonumber \\
&&
+4\,{
\frac { \left( s_{23}+s_{13} \right) ^{2}\epsilon}{s_{23}s_{13}}}+{\frac {-12\,{s_{12}}^{2}-12
\,m_H^4+8\,s_{12}m_H^2}{s_{23}s_{13}}}
+{\frac {4\,{s_{12}}^{2}+4\,m_H^4}{\epsilon\,s_{13}s_{23}}}
\Bigg) C_F\,{\rm Bub} \left( s_{12} \right) \nonumber \\
&&
+ \Bigg[  \Bigg( 8+4\,{
\frac { \left( s_{23}+s_{13} \right)  \left( s_{23}+m_H^2 \right) {\epsilon}^{2}}{
\left( s_{13}+s_{12} \right) s_{23}}}
-4\,{\frac { \left( s_{12}s_{23}+3\,s_{23}s_{13}-{s_{23}}^{2}+m_H^4
\right) \epsilon}{ \left( s_{13}+s_{12} \right) s_{23}}} \Bigg) N{C_F}^{2}\nonumber \\
&&
+\Bigg( 4\,{\frac { \left( s_{23}+s_{13} \right)  \left( 2\,s_{23}+s_{13} \right) {
\epsilon}^{2}}{s_{23}s_{13}}}-4\,{\frac { \left( s_{12}s_{13}-{s_{23}}^{2}-{s_{13}}^{2} \right)
\epsilon}{s_{23}s_{13}}}\nonumber \\
&&
-4\,{\frac {2\,{s_{12}}^{2}+{s_{23}}^{2}+s_{23}s_{13}+{s_{13}}^{2}+2\,m_H^4}{s_{23}s_{13}}}
+4\,{\frac {{s_{12}}^{2}+m_H^4}{\epsilon\,s_{13}s_{23}}} \Bigg)C_F\Bigg]
 {\rm Bub} \left( s_{23} \right) \nonumber \\
&&
+   \Bigg[\left( 8+4\,{\frac {
\left( s_{23}+s_{13} \right)  \left( s_{13}+m_H^2 \right) {\epsilon}^{2}}{ \left( s_{23}+s_{12}
\right) s_{13}}}-4\,{\frac { \left( s_{12}s_{13}+3\,s_{23}s_{13}-{s_{13}}^{2}+m_H^4 \right)
\epsilon}{ \left( s_{23}+s_{12} \right) s_{13}}} \right) N{C_F}^{2}\nonumber \\
&&
+ \Bigg( 4\,{
\frac { \left( s_{23}+2\,s_{13} \right)  \left( s_{23}+s_{13} \right) {\epsilon}^{2}}{s_{23}s_{13}}}
-4\,{\frac { \left( s_{12}s_{23}-{s_{23}}^{2}-{s_{13}}^{2} \right) \epsilon}{s_{23}s_{13}}}\nonumber \\
&&
-4\,{\frac {2\,{s_{12}}^{2}+{s_{23}}^{2}+s_{23}s_{13}+{s_{13}}^{2}+2\,m_H^4}{s_{23}s_{13}}}
+4\,{\frac {{s_{12}}^{
2}+m_H^4}{\epsilon\,s_{13}s_{23}}} \Bigg) C_F \Bigg] {\rm Bub} \left(
s_{13} \right)\nonumber \\
&&
+ \Bigg[  \Bigg( -8\,{\frac { \left( s_{23}+s_{13} \right) m_H^2 \left( 2
\,s_{23}s_{13}+s_{12}s_{23}+s_{12}s_{13} \right) {\epsilon}^{2}}{s_{13}s_{23} \left( s_{23}+s_{12} \right)  \left( s_{13}+s_{12}
\right) }}
-8\,{\frac {2\,{
s_{12}}^{2}+3\,s_{23}s_{13}+{s_{23}}^{2}+{s_{13}}^{2}+2\,m_H^4}{s_{23}s_{13}}}
\nonumber \\
&&
+{\frac { \left( 8\,{m_H^2}^{4}-8\,{s_{12}}^{2}{s_{13}}^{2}+16\,{s_{23}}^{2}{s_{13}
}^{2}-8\,{s_{23}}^{2}{s_{12}}^{2}+8\,{s_{12}}^{4}-8\,{s_{13}}^{3}m_H^2-8\,{s_{23}}^{3}m_H^2 \right)
\epsilon}{s_{13}s_{23} \left( s_{23}+s_{12} \right)  \left( s_{13}+s_{12} \right) }}\nonumber \\
&&
+8\,{\frac {{s_{12}}^{2}+{m_H^2}^{
2}}{\epsilon\,s_{13}s_{23}}} \Bigg) N{C_F}^{2}
+ \Bigg( -8\,{\frac {
\left( s_{23}+s_{13} \right) ^{2}{\epsilon}^{2}}{s_{23}s_{13}}}-4\,{\frac { \left( {s_{23}}^{2}+{s_{13}}^{2} \right) \epsilon}{s_{23}s_{13}}}\nonumber \\
&&
+4\,{\frac {2\,{s_{12}}^{2}+{s_{23}}^{2}+s_{23}s_{13}+{s_{13}}^
{2}+2\,m_H^4}{s_{23}s_{13}}}-4\,{\frac {{s_{12}}^{2}+m_H^4}{\epsilon\,s_{13}s_{23}}}
\Bigg) C_F \Bigg] {\rm Bub} \left( m_H^2 \right) \nonumber \\
&&
+ \left( 2\,{
\frac {s_{12}s_{13} \left( s_{23}+s_{13} \right) {\epsilon}^{2}}{s_{23}}}+2\,{\frac {s_{12} \left( s_{23}
+s_{13} \right) ^{2}\epsilon}{s_{23}}}-2\,{\frac {s_{12} \left( {s_{12}}^{2}+m_H^4
\right) }{s_{23}}} \right) C_F\,{\rm Box} \left( s_{12},s_{13},m_H^2  \right) \nonumber \\
&&
+\left( 2\,{\frac {s_{12}s_{23} \left( s_{23}+s_{13} \right) {\epsilon}^{2}}{s_{13}}}+2\,{
\frac {s_{12} \left( s_{23}+s_{13} \right) ^{2}\epsilon}{s_{13}}}-2\,{\frac {s_{12} \left( {s_{12}}^
{2}+m_H^4 \right) }{s_{13}}} \right) C_F\,{\rm Box} \left( s_{23},s_{12},m_H^2  \right) \nonumber \\
&&
+ \Big[  \left(  \left( -12\,s_{23}s_{13}-4\,{s_{23}}^{2}-4\,{s_{13}}^{2}
\right) \epsilon+4\,{s_{12}}^{2}+4\,m_H^4 \right) N{C_F}^{2}\nonumber \\
&&
+ \left(  \left( -6\,s_{23}s_{13}-2\,{s_{23}}^{2}-2\,{s_{13}}^{2} \right) \epsilon+2\,{s_{12}}^{2}+2\,m_H^4 \right) C_F \Big] {\rm Box} \left( s_{13},s_{23},m_H^2 \right) \Bigg\}\nonumber\\
&&
-\frac{1}{\epsilon}\left( \frac{3C_F}{2}+\frac{11C_A}{12} -\frac{T_Fn_f}{3} \right) \mathcal{A}_{H\rightarrow b\bar bg}^{(0)}.
\end{eqnarray}

There are only two different master integrals which appear here.
The bubble 
\beq
{\rm Bub} (s_{23})=\int \frac{d^dk}{i\pi^{\frac{d}{2}}} \frac{1}{k^2  (k+p_{23})^2 }=\frac{\Gamma(1+\epsilon) \Gamma(1-\epsilon)^2}{\epsilon\Gamma(2-2\epsilon)} (-s_{23})^{-\epsilon}
\eeq
as well as the following box integral
\beq
{\rm Box} (s_{23},s_{34}, m_H^2)=\int \frac{d^dk}{i\pi^{\frac{d}{2}}} \frac{1}{k^2 (k+p_2)^2 (k+p_{23})^2 (k+p_{234})^2}.
\eeq
which can be  expressed to all orders in $\epsilon$ in terms of hypergeometric functions
\begin{eqnarray}
{\rm Box} \left(s,t, M^2\right) &=& \frac{2\Gamma(1+\epsilon)\Gamma(1-\epsilon)^2}{\epsilon^2\Gamma(1-2\epsilon)} \frac{1}{st}  \Bigg[ -(-M^2)^{-\epsilon} {}_2F_1(1,-\epsilon,1-\epsilon,-\frac{uM^2}{st}) \nonumber \\
&+& (-t)^{-\epsilon} {}_2F_1(1,-\epsilon,1-\epsilon,-\frac{u}{s}) +(-s)^{-\epsilon} {}_2F_1(1,-\epsilon,1-\epsilon,-\frac{u}{t}) \Bigg] 
\end{eqnarray}
with  $u=M^2-s-t$.

\subsection{Decay to four partons}
For the decay to four partons we need 
\beq
H(p_H)\rightarrow b(p_1)+\bar b(p_2)+i(p_3)+j(p_4),\quad (ij)\in\{(q\bar q),(b\bar b),(gg)\}
\eeq
at tree-level
\beq
\left|\mathcal{M}_{H\rightarrow b\bar bij} \right|^2= y_b(\mu)^2\left(\frac{\alpha_s(\mu)}{\pi}\right)^2 \mathcal{A}_{H\rightarrow b\bar bij}^{(0)}+\mathcal{O}(\alpha_s(\mu)^3).
\eeq 
The $H\rightarrow b\bar{b}q\bar q$ amplitude can be expressed as 
\beq
\mathcal{A}_{H\rightarrow b\bar bq\bar q}^{(0)}=16\pi^4 \left( \frac{\mu^2 e^{\gamma_E}}{4\pi} \right)^{2\epsilon} \left[ 2NC_F A(p_1,p_2,p_3,p_4)\right]
\eeq
while for two $b\bar b$ pairs in the final state we get 
\begin{eqnarray}
\mathcal{A}_{H\rightarrow b\bar b b \bar b}^{(0)}&=& 16\pi^4 \left( \frac{\mu^2 e^{\gamma_E}}{4\pi} \right)^{2\epsilon} \times  \\
& & \Bigg\{ 2NC_F \left[ A(p_1,p_2,p_3,p_4) + A(p_1,p_4,p_3,p_2) + A(p_3,p_2,p_1,p_4) + A(p_3,p_4,p_1,p_2) \right]\nonumber \\
& & +2C_F \left[ B(p_1,p_2,p_3,p_4) +B(p_3,p_4,p_1,p_2)+B(p_4,p_3,p_2,p_1)+B(p_2,p_1,p_4,p_3) \right]\Bigg\} \nonumber
\end{eqnarray}
where
\begin{eqnarray}
A(p_1,p_2,p_3,p_4) &=& 2\bigg[ \frac{2(1-\epsilon)}{s_{34}}  -(1+\epsilon)\left(\frac{1}{s_{134}}+\frac{1}{s_{234}} \right)  
                              +m^2 (\epsilon-1)\left(  \frac{1}{s_{134}^2} +\frac{1}{s_{234}^2}        \right) \nonumber \\
			 & & +  \frac{s_{13}-s_{14}-2(s_{24}+m^2)-\epsilon(s_{13}+s_{14})}{s_{234}s_{34}}  
                             -\frac{2m^2}{s_{34}^2}\left( \frac{s_{14}}{s_{134}}-\frac{s_{24}}{s_{234}}\right)^2    \\ 
                         & & + \frac{s_{23}-s_{24}-2(s_{14}+m^2)-\epsilon(s_{23}+s_{24})}{s_{134}s_{34}}   
			     -\frac{2m^2}{s_{34}} \left( \frac{s_{14}}{s_{134}^2}+\frac{s_{24}}{s_{234}^2}\right) \nonumber \\
			 & & +2\frac{m^4+m^2(s_{14}+s_{24})+(s_{14}+s_{24})^2}{s_{134}s_{34}s_{234}}  
			     +2\frac{(1+\epsilon)m^2+s_{34}+2(s_{14}+s_{24})}{s_{134}s_{234}}\bigg]. \nonumber
\end{eqnarray}
The form factor $B(p_1,p_2,p_3,p_4)$ is presented in  Appendix \ref{RR}. The squared amplitude corresponding to the Higgs decaying into a $b\bar b$ plus two extra gluons 
can be expressed as
\beq
\mathcal{A}_{H\rightarrow b\bar {b} g  g}^{(0)}= 16\pi^4 \left( \frac{\mu^2 e^{\gamma_E}}{4\pi} \right)^{2\epsilon} \left\{ C_F A_{Hb\bar{b}gg} + NC_F^2 B_{Hb\bar{b}gg}+ C_F(1+2NC_F)  C_{Hb\bar{b}gg}  \right\}
\eeq
where 
\begin{eqnarray}
A_{Hb\bar{b}gg} &=& A^{(0)}_{Hb\bar{b}gg}+\epsilon A^{(1)}_{Hb\bar{b}gg}+\epsilon^2 A^{(2)}_{Hb\bar{b}gg} \nonumber \\
B_{Hb\bar{b}gg} &=& B^{(0)}_{Hb\bar{b}gg}+\epsilon B^{(1)}_{Hb\bar{b}gg}+\epsilon^2 B^{(2)}_{Hb\bar{b}gg} \nonumber \\
C_{Hb\bar{b}gg} &=& C^{(0)}_{Hb\bar{b}gg}+\epsilon C^{(1)}_{Hb\bar{b}gg}. 
\end{eqnarray}
The form factors $A^{(i)}_{Hb\bar{b}gg},B^{(i)}_{Hb\bar{b}gg}$ and $C^{(i)}_{Hb\bar{b}gg}$ can be found in Appendix \ref{RR}.

\section{Integration over phase space}
\label{sec:phase space}
The  matrix elements of Section~\ref{sec:matrix elements} are integrated over the phase space of the 
final-state partons. For convenience,  we  perform our calculation in the rest frame  of the Higgs  boson. Our results can 
be trivially extended to any frame of reference. The matrix-element for the $1 \to 2$ process $H \to b \bar{b}$ is independent 
of any phase space integration variables, and it is multiplied by the phase space volume:
\begin{equation}
\Phi_2^d=\int d\Phi_2^d = \frac{1}{(2\pi)^{d-2}} \frac{\Omega_{d-1}}{2^{d-1}}(m_H^2)^{-\epsilon}=\frac{1}{8\pi} \frac{(m_H^2)^{-\epsilon}(4\pi)^\epsilon \Gamma(1-\epsilon)}{\Gamma(2-2\epsilon)}.
\end{equation}

The matrix elements  for the $1\to 3$ processes  are integrated over the corresponding phase space with volume:
\begin{equation}
\Phi_3=\frac{\Phi_2(m_H^2)^{1-\epsilon}}{(4\pi)^{d/2}\Gamma(1-\epsilon)}  \int_0^1 d\lambda_1 d\lambda_2 (\lambda_1 \bar\lambda_1 \lambda_2)^{-\epsilon} \bar{\lambda}_2^{1-2\epsilon}.
\end{equation}
The invariants take the simple form:
\begin{eqnarray}
s_{12} & = & m_H^2 \lambda_2 \nonumber \\
s_{13} & = & m_H^2 \bar \lambda_2 \lambda_1\nonumber \\
s_{23} & = & m_H^2 \bar \lambda_2 \bar \lambda_1
\end{eqnarray}
In the above,  we have introduced the  shorthand notation
\begin{equation}
\bar \lambda \equiv 1-\lambda.
\end{equation}

We parameterize the phase space for the $1\to 4$ processes as follows: 
\beq
\Phi_4=  N_4 \int_0^1 d\lambda_1 d\lambda_2 d\lambda_3 d\lambda_4 d\lambda_5 [\lambda_1\bar{\lambda}_1\bar{\lambda_2}]^{1-2\epsilon}[\lambda_2\lambda_3\bar{\lambda}_3\lambda_4\bar{\lambda}_4]^{-\epsilon}\left[\sin(\lambda_5\pi)\right]^{-2\epsilon}
\eeq
where 
\beq
N_4=\frac { (m_H^2)^{2-3\epsilon} 2^{-13+8\epsilon} {\pi }^{-4+
3\,\epsilon}\left( 1-2\,\epsilon \right) }{  \Gamma  \left( 3/2-\epsilon
 \right)^{2}\Gamma  \left( 1-\epsilon \right) }.
\eeq
The invariants in this parameterization are
\begin{eqnarray}
s_{234} &=&m_H^2\lambda_1 \nonumber \\
s_{34}  &=&m_H^2\lambda_1\lambda_2 \nonumber \\
s_{23}  &=&m_H^2\lambda_1\bar{\lambda}_2 \lambda_4\nonumber \\
s_{24}  &=&m_H^2\lambda_1\bar{\lambda}_2 \bar{\lambda}_4\nonumber \\
s_{12}  &=&m_H^2\bar{\lambda}_1\bar{\lambda}_2 \bar{\lambda}_3\nonumber \\
s_{134} &=&m_H^2(\lambda_2+\lambda_3\bar{\lambda}_1\bar{\lambda}_2) 
\end{eqnarray}
and
\begin{eqnarray}
s_{13}  &=&m_H^2\bar{\lambda}_1 \left[ \lambda_4 \lambda_3+\lambda_2\bar{\lambda}_3\bar{\lambda}_4+2\cos(\lambda_5\pi)\sqrt{\lambda_2 \lambda_3 \bar{\lambda}_3 \lambda_4 \bar{\lambda}_4}     \right] \nonumber \\
s_{14}  &=&m_H^2\bar{\lambda}_1 \left[ \lambda_3 \bar{\lambda}_4+\lambda_2\bar{\lambda}_3\lambda_4-2\cos(\lambda_5\pi)\sqrt{\lambda_2 \lambda_3 \bar{\lambda}_3 \lambda_4 \bar{\lambda}_4}     \right].  
\end{eqnarray}
This parameterization is  very similar to the  ones used  in Refs~\cite{GehrmannDeRidder:2003bm,Anastasiou:2003gr}.  

In the following  sections  we  shall describe the method of non-linear mappings that we have  employed  in order to perform the non-trivial 
phase space integrations  over the matrix elements at NNLO.  Specifically, we  shall discuss the ``real-virtual'' contributions which require both 
an integration over the  three-parton phase space and  an one-loop amplitude, and the ``double-real''  contributions  which require the  integration  of tree matrix elements over the four-parton phase space.  

\section{Real-Virtual NNLO contribution}
\label{sec:real-virtual}

In the one loop amplitude for $H\rightarrow b\bar b g$ we will face integrals like 
\beq
\int d\Phi_3 \frac{{}_2F_1(1;-\epsilon,1-\epsilon,-\frac{s_{23}}{s_{13}} )}{s_{23}s_{13}}
\eeq
Since the numerator in this case is not well defined for $s_{23},s_{13} \rightarrow 0$, we can not simply apply a naive 
 ``plus-prescription''  subtraction, 
\begin{equation}
\lambda^{-1+a\epsilon}=\frac{\delta(\lambda)}{a\epsilon}+\sum_{n=0}^\infty \frac{(a\epsilon)^n}{n!}\left[\frac{\log^n(\lambda)}{\lambda}\right]_+ 
\end{equation}
to extract the singularities. To tackle this difficulty we use Euler's integral representation of the hypergeometric function, and apply a non-linear 
mapping to disentangle an overlapping singularity of the integration variable in the hypergeometric  representation 
and the phase space variables which generate $s_{13}, s_{23}$.
Defining 
\begin{equation}
F(z)=\int_0^1 dx_3 \frac{x_3^{-1-\epsilon}}{1+zx_3} =\frac{  {}_2F_1(1,-\epsilon,1-\epsilon,-z)}{-\epsilon}
\end{equation}
we apply the transformation,  
\begin{equation}
\label{eq:rv-mapping}
x_3 \mapsto \beta(x_3,1,z),\quad \quad \beta(x,A,B) := \frac{xA}{A+\bar{x}B}, 
\end{equation}
which yields
\beq
F(z)=(1+z)^\epsilon \int_0^1 dx_3 x_3^{-1-\epsilon}\left(1- \frac{x_3z}{1+z} \right)^\epsilon .
\eeq
This  representation of the hypergeometric function has finite values  at the points where we need  to make  a
subtraction, 
\begin{eqnarray}
{}_2F_1(-\epsilon,-\epsilon,1-\epsilon,0) &=&1 \nonumber \\
{}_2F_1(-\epsilon,-\epsilon,1-\epsilon,1) &=& \Gamma(1+\epsilon)\Gamma(1-\epsilon),  
\end{eqnarray}
and we can obtain the Laurent expansion in $\epsilon$ of the real-virtual contribution easily with a 
``plus-prescription'' subtraction. For non-special values  of the kinematic variables the $\epsilon-$expansion 
of  hypergeometric function reads: 
\beq
 {}_2F_1\left(-\epsilon,-\epsilon;1-\epsilon;z\right)
= \left(1-z\right)^{\epsilon} \left(1 - \epsilon \log\left(1-z\right)
-\sum_{k=2}^\infty \epsilon^k {\rm
Li}_k\left(\frac{z}{z-1}\right)\right).
\eeq

The mapping of Eq.~\ref{eq:rv-mapping} simply re-derives a well known identity
\beq
{}_2F_1(a,b,c;z)=(1-z)^{-b} {}_2F_1(c-a,b,c;\frac{z}{z-1}),
\eeq	
and there would be no need  of it  had we started  with the following representation of the box master integral:
\begin{eqnarray}
{\rm Box} (u,t,M^2) &=& \frac{2\Gamma(1+\epsilon)\Gamma(1-\epsilon)^2}{\epsilon^2\Gamma(1-2\epsilon)} (ut)^{-1-\epsilon} \nonumber \\
&\times& \Bigg[ -(-M^2)^{-\epsilon} (ut+sM^2)^\epsilon {}_2F_1(-\epsilon,-\epsilon,1-\epsilon,\frac{sM^2}{ut+sM^2})  \\
&+& (-u-s)^\epsilon {}_2F_1(-\epsilon,-\epsilon,1-\epsilon,\frac{s}{u+s}) + (-t-s)^\epsilon {}_2F_1(-\epsilon,-\epsilon,1-\epsilon,\frac{s}{t+s}) \Bigg] \nonumber
\end{eqnarray}
On the other hand,  more  complicated master integrals for  one-loop amplitudes  do not always have  representations  in terms  
of  hypergeometric  functions with known expansions in $\epsilon$ and variable limits.   However, it  is always possible to  derive   
a Feynman representation or a hypergeometric representation (via a Mellin-Barnes representation) for them and attempt a direct numerical 
evaluation of the combined phase space and virtual multiple integral using non-linear transformations to factorize  the  singularities.  
We have compared the two approaches: 
\begin{enumerate}
\item Evaluate numerically the coefficients of the Laurent expansion for the two-dimensional phase space integral after we express the  
loop amplitude  and  its  hypergeometric functions in terms  of polylogarithms,   
\item  Evaluate numerically the coefficients of the Laurent expansion for the three-dimensional combined 
phase space and loop integral.  
\end{enumerate}
We  find that  the three-dimensional integration is as fast as the two-dimensional integration where we  require the evaluation 
of polylogarithmic functions.

\section{Integration of double-real contributions} 
\label{sec:double-real}
phase space  integrals for four-parton processes develop singularities at multiple soft and  collinear 
kinematic configurations. These singularities are overlapping.   In this section, we demonstrate how they can be factorized 
with non-linear transformations. We begin the discussion with an integral with four singular denominators, 
\begin{equation}
\label{eq:RRint1}
I_1[J]=\int d\Phi_4 \frac{J(p_1,p_2,p_3,p_4)}{s_{12}s_{34}s_{123}s_{234}}.
\end{equation}
where the function $J$ represents  a non singular function of the partonic momenta, composed  of numerators  in the 
squared matrix elements and the infrared-safe observable ${\cal J}$.
We disentangle the overlapping singularities using
\begin{equation}
\bar \lambda_3\mapsto \alpha(\bar \lambda_3,\lambda_4,1), 
\end{equation}
where 
\begin{equation}
\alpha(x,A,B) := \frac{xA}{xA+\bar{x}B}.
\end{equation}
We note that $x \mapsto \alpha(x,A,B) \Rightarrow \bar{x} \mapsto \alpha(\bar{x},B,A)$.
After this transformation, the  expansion in $\epsilon$ is straightforward with simple subtractions. For the special case $J=1$, we obtain
\begin{equation}
I_1\left[J=1\right]=N_4\left[61.76(2)+\frac{8.554(2)}{\epsilon}- \frac{0.1710(2)}{\epsilon^2}+\frac{0.34657(2)}{\epsilon^3} +\frac{0.25}{\epsilon^4}\right].
\end{equation}

Other integrals may be mapped to $I_1$ of  Eq.~\ref{eq:RRint1} with a relabeling of the partonic momenta. 
For  example, the integral
\begin{equation}
I_1=\int d\Phi_4 \frac{J(p_1,p_2,p_3,p_4)}{s_{14}s_{23}s_{134}s_{234}}, 
\end{equation}
which has an apparently more complicated singularity structure due  to $s_{14}$ in the 
denominator,  is transformed to the integral of Eq.~\ref{eq:RRint1} by exchanging $p_2$ and $p_4$. 

A second  class of  integrals has the following  denominator structure:
\begin{eqnarray}
\label{eq:RRint2}
I_2[J] &=&\int d\Phi_4 \frac{J(p_1,p_2,p_3,p_4)}{s_{13}s_{23}s_{134}s_{234}} 
\end{eqnarray}
Notice, that the integrand is singular not only at the boundaries of the integration variables 
$\lambda_i$, but  it also develops a ``line singularity''  inside  the integration volume.  We  eliminate this by 
casting the integral in the form
\begin{eqnarray}
\label{eq:RRint2_a}
I_2[J]    &=&\int d\Phi_4 \frac{(J(p_1,p_2,p_3,p_4)+J(p_2,p_1,p_3,p_4))s_{24}}{s_{23}s_{134}s_{234}(s_{13} s_{24}+s_{14} s_{23})}. 
\end{eqnarray}
It  is easy to see  that  Eq.~\ref{eq:RRint2_a} is  equivalent  to $I_2[J]$, by exchanging  the momenta $p_1 \leftrightarrow p_2$ in the term 
with $J(p_2,p_1,p_3,p_4)$.  
The denominator
\beq
s_{13} s_{24}+s_{14} s_{23}=m_H^4\lambda_1\bar\lambda_1\bar\lambda_2 \left[ 2\lambda_3 \lambda_4\bar\lambda_4+\lambda_2\bar\lambda_3(\lambda_4^2+\bar\lambda_4^2)+2\cos(\pi\lambda_5)(1-2\lambda_4)\sqrt{\lambda_2\lambda_3\bar\lambda_3\lambda_4\bar\lambda_4}\right]
\eeq
is only singular at  the boundaries of the integration region.  Applying the following sequence of mappings
\begin{eqnarray}
\lambda_2  &\mapsto& \alpha(\lambda_2,\lambda_3,1) \nonumber \\
\lambda_4  &\mapsto& \alpha(\lambda_4,\lambda_2 \bar \lambda_3,1) \nonumber \\
\lambda_2  &\mapsto& \alpha(\lambda_2,\bar \lambda_1,1)  
\end{eqnarray}
leads to a factorized form. Numerically we then obtain
\beq
I_2\left[ J =1 \right]=N_4\left[ -201.16(3)-\frac{68.426(4)}{\epsilon}- \frac{12.027(8)}{\epsilon^2}+\frac{1.0397(1)}{\epsilon^3} +\frac{0.75}{\epsilon^4}\right].
\eeq

A last class  of singular integrals with four denominators is:
\begin{eqnarray}
I_3\left[ J \right] &=&\int d\Phi_4 \frac{J(p_1,p_2,p_3,p_4)}{s_{13}s_{23}s_{14}s_{24}} \nonumber \\
    &=&\int d\Phi_4 \frac{J(p_1,p_3,p_2,p_4)}{s_{12}s_{34}s_{14}s_{23}} \nonumber \\
    &=&\int d\Phi_4 (J(p_1,p_3,p_2,p_4)+J(p_2,p_4,p_1,p_3))\frac{s_{24}}{s_{34}s_{12}s_{23}(s_{13} s_{23}+s_{14} s_{24})}. 
\end{eqnarray}
The term in parenthesis in the last denominator is given by
\beq
s_{13} s_{23}+s_{14} s_{24}=m_H^4\lambda_1\bar\lambda_1\bar\lambda_2 \left[ 2\lambda_2\bar\lambda_3 \lambda_4\bar\lambda_4+\lambda_3(\lambda_4^2+\bar\lambda_4^2)+2\cos(\pi\lambda_5)(2\lambda_4-1)\sqrt{\lambda_2\lambda_3\bar\lambda_3\lambda	_4\bar\lambda_4}\right]
\eeq
We let $dI_3=\lambda_2dI_3+\bar \lambda_2 dI_3$ and apply 
\begin{eqnarray}
\lambda_2  &\mapsto& \alpha(\lambda_2,\lambda_3,1) \nonumber \\
\lambda_{3,4} &\mapsto& \alpha(\lambda_{3,4},\bar\lambda_2,1)  
\end{eqnarray}
to  $\bar \lambda_2 dI_3$ and
\begin{equation}
\lambda_4\mapsto \alpha(\lambda_4,\lambda_3,1)
\end{equation}
to $ \lambda_2 dI_3$. After these  transformations, both integrals have only factorized singularities  and  can be 
evaluated  numerically.  We obtain:
\begin{equation}
I_3\left[ J =1 \right]=N_4\left[ -292.54(4)-\frac{217.030(9)}{\epsilon}- \frac{52.768(2)}{\epsilon^2}+\frac{6.9314(3)}{\epsilon^3} +\frac{5}{\epsilon^4}\right].
\end{equation}

Other integrals with four denominators are mapped to  $I_{1,2,3}[J]$ with simple relabeling  of the partonic momenta.  
Integrals with fewer denominators  are simpler and  the non-linear mappings for them are obvious upon inspection, following the 
examples of  Ref.~\cite{Anastasiou:2010pw}. One case which requires special attention arises when a denominator is raised  to the second power 
in the squares of  diagrams where a gluon splits into a $q\bar q$ or $gg$ pair as has already been 
discussed in the literature (see, for example, \cite{Anastasiou:2005qj,Biswas:2009rb}). These quadratic  singularities are fake.   
A combination of terms, such as  
\beq
\frac{2m_H^2}{s_{34}^2}\left( \frac{s_{14}}{s_{134}}-\frac{s_{24}}{s_{234}}\right)^2
\eeq
is  the one  which emerges  in  the physical matrix elements. After a non-linear mapping 
\begin{equation}
\lambda_2 \to \alpha(\lambda_2, \lambda_3, 1)
\end{equation}
to factorize an overlapping singularity, the   integrand is also free of quadratic singularities in the integration variables $\lambda_i$.

\section{Numerical results}
\label{sec:numerics}

In this section we present our numerical results for the decay rate $\Gamma_{H \to b \bar{b} X}[{\cal J}]$ for selected infrared safe 
observables ${\cal J}$. We select the value  of the renormalization scale $\mu = m_H$. First, we  compute  the  inclusive decay width. Our numerical result is, 
\begin{equation}
\Gamma_{H\rightarrow b\bar b}^{NNLO}=\Gamma_{H\rightarrow b\bar b}^{LO}\left[ 1 + \left(\frac{\alpha_s}{\pi} \right) 5.6666(4)       + \left(\frac{\alpha_s}{\pi} \right)^2   29.14(2) +\mathcal{O}(\alpha_s^3) \right]  
\end{equation}
with 
\begin{equation}
\Gamma_{H\rightarrow b\bar b}^{LO}=\frac{y_b^2m_HN}{8\pi}.
\end{equation}
This is in agreement with the known analytic expression~\cite{{Baikov:2005rw}}
\begin{equation}
\Gamma_{H\rightarrow b\bar b}^{NNLO}=\Gamma_{H\rightarrow b\bar b}^{LO}\left[ 1 + \left(\frac{\alpha_s}{\pi} \right) 5.6666666..       + \left(\frac{\alpha_s}{\pi} \right)^2   29.146714.. +\mathcal{O}(\alpha_s^3) \right].  
\end{equation}
With our numerical program, we can compute the  decay rate  for arbitrary infrared-safe  observables.  
As an example  we present our results for the 2,3 and 4 jet rates 
using the JADE algorithm~\cite{Bartel:1986ua} with parameter $y_{cut}=0.01$, specifically we cluster particles using the distance measure $y_{ij}=(p_i+p_j)^2$ and recombine them by adding their momenta:
\begin{eqnarray}
\Gamma_{H\rightarrow b\bar b}^{LO}(4 {\rm Jet Rate})   &=& \Gamma_{H\rightarrow b\bar b}^{LO}\left[  + \left(\frac{\alpha_s}{\pi} \right)^2 94.1(1)    +\mathcal{O}(\alpha_s^3) \right]  \nonumber \\
\Gamma_{H\rightarrow b\bar b}^{NLO}(3 {\rm Jet Rate})  &=& \Gamma_{H\rightarrow b\bar b}^{LO}\left[  + \left(\frac{\alpha_s}{\pi} \right) 19.258(4)       + \left(\frac{\alpha_s}{\pi} \right)^2  241(2)   +\mathcal{O}(\alpha_s^3) \right]   \nonumber \\
\Gamma_{H\rightarrow b\bar b}^{NNLO}(2 {\rm Jet Rate}) &=& \Gamma_{H\rightarrow b\bar b}^{LO}\left[ 1 - \left(\frac{\alpha_s}{\pi} \right)13.591(6)        - \left(\frac{\alpha_s}{\pi} \right)^2 307(2)    +\mathcal{O}(\alpha_s^3) \right].    
\end{eqnarray}
We have checked that  our  results  for the jet-rates add up to the inclusive rate for general $y_{cut}$ values. 
 
\begin{figure}
\begin{center}
\includegraphics[scale=1]{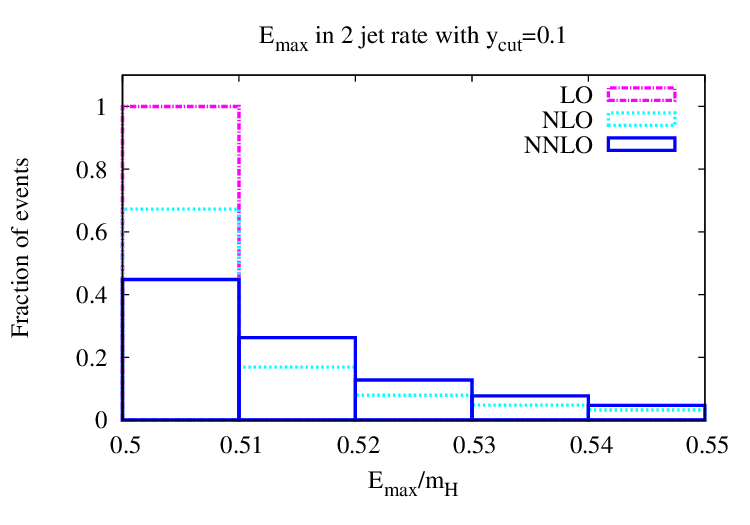}
\end{center}
\caption{\label{fig:Emax}  The energy spectrum of the leading jet  in the decay $H \to b \bar{b}$ in the rest frame of the Higgs boson through NNLO. The 
jet clustering is performed with the JADE algorithm with $y_{cut}=0.1$.}
\end{figure}

In Fig.~\ref{fig:Emax}, we present  the leading jet energy, $E_{max}$, in the rest frame of the Higgs boson, for events with two jets 
($y_{cut}=0.1$).   This  is  a new result which can only be obtained  with a fully differential NNLO calculation and cannot  be inferred from 
the inclusive decay width and NLO calculations. At leading order, $E_{max} = \frac{m_H}{2}$. At higher orders, there is a range of jet energies 
which are kinematically allowed.  
We  choose  a value of $\alpha_s(m_Z) = 0.118$ at the Z boson mass and evolve consistently through LO, NLO and NNLO up to the Higgs  boson mass, 
which we  assume to be $m_H = 120 {\rm GeV}$. 

The above  numerical results demonstrate  the applicability of our method  to physical processes. 
A number of phenomenological studies  which are relevant to the searches  of the Higgs boson can be  made.  
It is  easy to interface our  numerical code  with  a Monte-Carlo code for the  production of  a Higgs boson. 
We we will present complete phenomenological studies pursuing this direction in the future.


\section{Conclusions}
\label{sec:conclusions}

In this paper, we  present a first physical application of a new
method for the factorization of overlapping singularities in
phase space and loop integrations.  We compute the fully differential
decay width of a Higgs boson to a bottom-quark pair. We produce the
required tree, one-loop and two-loop amplitudes with standard
Feynman diagrammatic methods. 

Our article focuses on the phase space
integrations which emerge at NNLO.  We apply non-linear mappings to
factorize all overlapping singularities in all
real-virtual and double-real integrations. Consequently, we 
perform the expansion of all  integrals in the dimension
regulator $\epsilon$ with simple subtractions. The formalism allows
for the computation of the decay rate for arbitrary physical
observables. 

We verify that we can reproduce the known results for the NNLO
inclusive decay width and compute the differential two,three and four
jet rates with the JADE algorithm. We also present the leading jet
energy distributions, a new result that cannot be inferred from
previous calculations.  In the future, we will interface our NNLO
Monte-Carlo  to Monte-Carlo programs for the production of a Higgs
boson.  

We believe that our method for NNLO computations is powerful and
suitable for complicated physical processes. We are looking forward to 
further apply our method on other interesting processes at  hadron
colliders.

\section{Acknowledgements}
We thank Andrea Banfi and Zoltan Kunszt for useful discussions. 
This research is supported by the ERC Starting Grant for the project
``IterQCD'' and the Swiss National Foundation under contract SNF
200020-126632.


\begin{appendix}
 
\section{Double-real matrix elements}
\label{RR}
The function $B(p_1,p_2,p_3,p_4)$ occurring in the $H\rightarrow b\bar b b \bar b$ squared amplitude 
can be expressed as
\begin{eqnarray}
B(p_1,p_2,p_3,p_4) &=&2\,{s_{{34}}}^{-1}-2\,{s_{{134}}}^{-1}-4\,{s_{{234}}}^{-1}+2\,{s_{{124
}}}^{-1}-2\,{\frac {s_{{12}} \left( {s_{{13}}}^{2}+2\,{s_{{12}}}^{2}+2
\,s_{{12}}s_{{13}} \right) }{s_{{14}}s_{{34}}s_{{124}}s_{{234}}}}\nonumber \\
& &
+4\,{
\frac {s_{{12}}s_{{23}}}{s_{{34}}s_{{134}}s_{{234}}}}-2\,{\frac {s_{{
12}} \left( s_{{23}}+2\,s_{{13}}+s_{{12}} \right) }{s_{{14}}s_{{34}}s_
{{124}}}}+4\,{\frac {s_{{12}}s_{{23}}}{s_{{14}}s_{{124}}s_{{134}}}}+\nonumber \\
& &
 \Bigg( {\frac {4\,s_{{13}}s_{{14}}+2\,{s_{{13}}}^{2}+4\,s_{{12}}s_{{
13}}+2\,{s_{{14}}}^{2}+2\,s_{{12}}s_{{14}}}{s_{{34}}s_{{124}}s_{{234}}
}}+{\frac {-2\,s_{{24}}-4\,s_{{13}}}{s_{{14}}s_{{34}}}}\nonumber \\
& &+{\frac {-2\,s_
{{24}}+2\,s_{{12}}-2\,s_{{23}}}{s_{{14}}s_{{234}}}}+{\frac {4\,s_{{14}
}-4\,s_{{23}}}{s_{{124}}s_{{134}}}}+{\frac {-4\,s_{{12}}-4\,s_{{14}}-4
\,s_{{13}}}{s_{{134}}s_{{234}}}}\nonumber \\
& &+{\frac {2\,s_{{14}}+4\,s_{{13}}+2\,s_
{{23}}}{s_{{34}}s_{{124}}}}+2\,{\frac {s_{{12}}s_{{13}}}{s_{{14}}s_{{
34}}s_{{124}}}}+{\frac {-2\,s_{{24}}+2\,s_{{12}}}{s_{{14}}s_{{124}}}}-
2\,{\frac {s_{{13}} \left( -s_{{23}}+s_{{13}} \right) }{s_{{14}}s_{{34
}}s_{{234}}}}\nonumber \\
& &+{\frac {-4\,s_{{13}}-2\,s_{{14}}+2\,s_{{23}}}{s_{{34}}s_
{{234}}}}+{\frac {4\,s_{{12}}s_{{13}}+2\,s_{{24}}s_{{23}}+2\,{s_{{24}}
}^{2}}{s_{{14}}s_{{124}}s_{{234}}}}+{\frac {-4\,s_{{12}}-4\,s_{{24}}-4
\,s_{{23}}}{{s_{{134}}}^{2}}}\nonumber \\
& &+{\frac {2\,s_{{24}}+2\,s_{{23}}}{s_{{34}
}s_{{134}}}}+2\,{s_{{14}}}^{-1}+2\,{\frac {s_{{12}}{s_{{13}}}^{2}}{s_{
{14}}s_{{34}}s_{{124}}s_{{234}}}}-8\,{s_{{134}}}^{-1}+2\,{s_{{234}}}^{
-1}+{\frac {2\,s_{{12}}+2\,s_{{24}}}{s_{{14}}s_{{134}}}}\nonumber \\
& &-2\,{s_{{124}}
}^{-1} \Bigg) {\epsilon}^{2}+ \Bigg( -4\,{s_{{14}}}^{-1}+6\,{s_{{34}}
}^{-1}+6\,{s_{{134}}}^{-1}+4\,{s_{{234}}}^{-1}-2\,{s_{{124}}}^{-1}\nonumber \\
& &+2\,
{\frac {s_{{12}}{s_{{13}}}^{2}}{s_{{14}}s_{{34}}s_{{124}}s_{{234}}}}-4
\,{\frac {s_{{12}}s_{{23}}}{s_{{34}}s_{{134}}s_{{234}}}}-4\,{\frac {s_
{{12}}s_{{23}}}{s_{{14}}s_{{124}}s_{{134}}}}\nonumber \\
& &-2\,{\frac {s_{{12}}
 \left( -s_{{23}}-2\,s_{{13}}+s_{{12}} \right) }{s_{{14}}s_{{34}}s_{{
124}}}}+{\frac {4\,s_{{12}}+4\,s_{{24}}+4\,s_{{23}}}{{s_{{134}}}^{2}}}\nonumber \\
& &
+{\frac {-6\,s_{{24}}-8\,s_{{13}}}{s_{{14}}s_{{34}}}}+{\frac {-6\,s_{{
12}}-2\,s_{{23}}+6\,s_{{34}}+8\,s_{{13}}}{s_{{14}}s_{{134}}}}+{\frac {
6\,s_{{13}}+6\,s_{{24}}-2\,s_{{14}}+2\,s_{{23}}}{s_{{34}}s_{{134}}}}\nonumber \\
& &+{
\frac {6\,s_{{13}}+4\,s_{{24}}+10\,s_{{12}}}{s_{{14}}s_{{234}}}}+{
\frac {-2\,s_{{23}}-4\,s_{{24}}+2\,s_{{14}}+4\,s_{{12}}}{s_{{34}}s_{{
234}}}}+{\frac {-4\,s_{{13}}-2\,s_{{14}}-4\,s_{{12}}}{s_{{134}}s_{{234
}}}}\nonumber \\
& &+{\frac {-2\,s_{{24}}-2\,s_{{12}}}{s_{{14}}s_{{124}}}}+{\frac {6\,
s_{{23}}-8\,s_{{12}}-8\,s_{{14}}+6\,s_{{13}}}{s_{{34}}s_{{124}}}}+{
\frac {-2\,s_{{13}}-4\,s_{{23}}+2\,s_{{14}}}{s_{{124}}s_{{134}}}}\nonumber \\
& &+{
\frac {-4\,s_{{14}}-4\,s_{{13}}-8\,s_{{12}}}{s_{{124}}s_{{234}}}}+4\,{
\frac {s_{{13}} \left( s_{{12}}+s_{{24}}+s_{{23}} \right) }{s_{{14}}{s
_{{134}}}^{2}}}+4\,{\frac {s_{{13}} \left( s_{{12}}+s_{{24}}+s_{{23}}
 \right) }{s_{{34}}{s_{{134}}}^{2}}}\nonumber \\
& &+{\frac {4\,s_{{13}}s_{{23}}+2\,s_
{{12}}s_{{23}}-2\,{s_{{13}}}^{2}-2\,{s_{{23}}}^{2}}{s_{{14}}s_{{34}}s_
{{234}}}}\nonumber \\
& &+{\frac {2\,{s_{{12}}}^{2}+4\,{s_{{23}}}^{2}+8\,s_{{24}}s_{{
23}}+4\,s_{{12}}s_{{23}}+6\,{s_{{24}}}^{2}+8\,s_{{12}}s_{{24}}}{s_{{14
}}s_{{234}}s_{{134}}}}\nonumber \\
& &+{\frac {-4\,s_{{12}}s_{{23}}+6\,{s_{{14}}}^{2}+
4\,s_{{12}}s_{{14}}+2\,{s_{{23}}}^{2}+2\,{s_{{12}}}^{2}-8\,s_{{23}}s_{
{14}}}{s_{{34}}s_{{124}}s_{{134}}}}+{\frac {2\,{s_{{24}}}^{2}+2\,s_{{
24}}s_{{23}}-4\,{s_{{12}}}^{2}}{s_{{14}}s_{{124}}s_{{234}}}}\nonumber \\
& &+{\frac {-
4\,{s_{{12}}}^{2}-6\,s_{{12}}s_{{14}}+2\,{s_{{13}}}^{2}-2\,{s_{{14}}}^
{2}}{s_{{34}}s_{{124}}s_{{234}}}} \Bigg) \epsilon+{\frac {4\,s_{{12}}
+4\,s_{{23}}+8\,s_{{13}}}{s_{{14}}s_{{34}}}}\nonumber \\
& &+{\frac {-6\,s_{{13}}-2\,s
_{{14}}-4\,s_{{24}}}{s_{{34}}s_{{134}}}}+{\frac {2\,s_{{23}}-2\,s_{{24
}}+4\,s_{{12}}-2\,s_{{34}}-4\,s_{{13}}}{s_{{14}}s_{{134}}}}\nonumber \\
& &+{\frac {-4
\,s_{{23}}-10\,s_{{12}}+4\,s_{{24}}-6\,s_{{13}}}{s_{{14}}s_{{234}}}}+{
\frac {4\,s_{{12}}+4\,s_{{13}}+2\,s_{{23}}+2\,s_{{14}}}{s_{{34}}s_{{
234}}}}\nonumber \\
& &+{\frac {2\,s_{{14}}+4\,s_{{13}}+4\,s_{{12}}}{s_{{134}}s_{{234}
}}}+{\frac {2\,s_{{12}}+2\,s_{{24}}}{s_{{14}}s_{{124}}}}+{\frac {-6\,s
_{{13}}-6\,s_{{23}}-4\,s_{{12}}}{s_{{34}}s_{{124}}}}+{\frac {2\,s_{{13
}}-2\,s_{{14}}+4\,s_{{23}}}{s_{{124}}s_{{134}}}}\nonumber \\
& &+{\frac {4\,s_{{14}}+4
\,s_{{13}}+8\,s_{{12}}}{s_{{124}}s_{{234}}}}-4\,{\frac {s_{{13}}
 \left( s_{{12}}+s_{{24}}+s_{{23}} \right) }{s_{{14}}{s_{{134}}}^{2}}}
-4\,{\frac {s_{{13}} \left( s_{{12}}+s_{{24}}+s_{{23}} \right) }{s_{{
34}}{s_{{134}}}^{2}}}\nonumber \\
& &+{\frac {-2\,{s_{{23}}}^{2}-2\,s_{{12}}s_{{23}}+2
\,{s_{{13}}}^{2}-4\,s_{{13}}s_{{23}}+4\,{s_{{12}}}^{2}+4\,{s_{{24}}}^{
2}+4\,s_{{12}}s_{{13}}-4\,s_{{24}}s_{{13}}-4\,s_{{12}}s_{{24}}}{s_{{14
}}s_{{34}}s_{{234}}}}\nonumber \\
& &+{\frac {-2\,{s_{{12}}}^{2}-2\,{s_{{24}}}^{2}-4\,
{s_{{23}}}^{2}-4\,s_{{12}}s_{{24}}-4\,s_{{24}}s_{{23}}-4\,s_{{12}}s_{{
23}}}{s_{{14}}s_{{234}}s_{{134}}}}\nonumber \\
& &+{\frac {-2\,{s_{{23}}}^{2}-2\,{s_{{
12}}}^{2}+4\,s_{{23}}s_{{14}}-2\,{s_{{14}}}^{2}}{s_{{34}}s_{{124}}s_{{
134}}}}+{\frac {4\,{s_{{12}}}^{2}-2\,s_{{24}}s_{{23}}-2\,{s_{{24}}}^{2
}}{s_{{14}}s_{{124}}s_{{234}}}}\nonumber \\
& &+{\frac {-4\,s_{{13}}s_{{14}}-8\,{s_{{
12}}}^{2}-2\,{s_{{13}}}^{2}-2\,{s_{{14}}}^{2}-8\,s_{{12}}s_{{13}}-6\,s
_{{12}}s_{{14}}}{s_{{34}}s_{{124}}s_{{234}}}}
\end{eqnarray}

\begin{eqnarray}
A^{(0)}_{Hb\bar{b}gg} &=&  20\,{s_{234}}^{-1}+20\,{s_{134}}^{-1}+32\,{s_{{13}}}^{-1}+24\,{s
_{{24}}}^{-1}+24\,{s_{{14}}}^{-1}+32\,{s_{{23}}}^{-1}\nonumber \\
&&
+4\,{\frac {3\,s_
{{13}}-6\,{m_{{H}}}^{2}+3\,s_{{14}}+6\,s_{{34}}}{s_{{24}}s_{{23}}}}+8
\,{\frac {3\,s_{{24}}-6\,{m_{{H}}}^{2}+3\,s_{{14}}+3\,s_{{34}}}{s_{{13
}}s_{{23}}}}\nonumber \\
&&
+4\,{\frac {6\,{m_{{H}}}^{2}-3\,s_{{13}}+3\,s_{{34}}+3\,s_
{{24}}}{s_{234}\,s_{{14}}}}+4\,{\frac {-s_{{24}}-s_{{14}}+s_{{34}}+
4\,{m_{{H}}}^{2}}{s_{234}\,s_{{13}}}}\nonumber \\
&&
-4\,{\frac {-3\,s_{{14}}-6\,{m
_{{H}}}^{2}+3\,s_{{23}}-3\,s_{{34}}}{s_{134}\,s_{{24}}}}+4\,{\frac 
{3\,s_{{23}}+6\,s_{{34}}-6\,{m_{{H}}}^{2}+3\,s_{{24}}}{s_{{13}}s_{{14}
}}}\nonumber \\
&&
+8\,{\frac {-5\,s_{{34}}-8\,{m_{{H}}}^{2}-4\,s_{{24}}-4\,s_{{14}}}{
s_{134}\,s_{234}}}+8\,{\frac {{m_{{H}}}^{2}}{{s_{134}}^{2}}}+
8\,{\frac {{m_{{H}}}^{2}}{{s_{234}}^{2}}}\nonumber \\
&&
-4\,{\frac {8\,{m_{{H}}}^{2}-5\,s_{{14}}-7\,s_{{34}}-5\,s_{{23}}}{s_{{24}}s_{{13}}}}+4\,{\frac {
-s_{{24}}-s_{{14}}+s_{{34}}+4\,{m_{{H}}}^{2}}{s_{134}\,s_{{23}}}}\nonumber \\
&&
-4\,{\frac {-5\,s_{{24}}+8\,{m_{{H}}}^{2}
-5\,s_{{13}}-7\,s_{{34}}}{s_{{14}}s_{{23}}}}-4\,{\frac {-s_{{14}}-2\,s_{{34}}+s_{{13}}-3\,{m_{{H}}}^
{2}}{s_{234}\,s_{{24}}}}\nonumber \\
&&
+8\,{\frac {3\,s_{{23}}+3\,s_{{13}}+3\,s_{{
34}}-6\,{m_{{H}}}^{2}}{s_{{24}}s_{{14}}}}-4\,{\frac {-3\,{m_{{H}}}^{2}
-2\,s_{{34}}-s_{{24}}+s_{{23}}}{s_{134}\,s_{{14}}}}\nonumber \\
&&
+4\,{\frac {
 \left( {m_{{H}}}^{2}+s_{{34}} \right)  \left( -2\,{m_{{H}}}^{4}-{s_{{
34}}}^{2}-2\,{m_{{H}}}^{2}s_{{34}} \right) }{s_{234}\,s_{134}\,s
_{{13}}s_{{23}}}}\nonumber \\
&&
+4\,{\frac { \left( {m_{{H}}}^{2}+s_{{34}} \right) 
 \left( -2\,{m_{{H}}}^{4}-{s_{{34}}}^{2}-2\,{m_{{H}}}^{2}s_{{34}}
 \right) }{s_{234}\,s_{134}\,s_{{14}}s_{{24}}}}\nonumber \\
&&
 +4\,{\frac {
 \left( -s_{{34}}+{m_{{H}}}^{2} \right)  \left( -2\,{m_{{H}}}^{4}-{s_{
{34}}}^{2}+2\,{m_{{H}}}^{2}s_{{34}} \right) }{s_{{13}}s_{{14}}s_{{23}}
s_{{24}}}}\nonumber \\
&&
-4\,{\frac {3\,{m_{{H}}}^{2}s_{{13}}-{s_{{34}}}^{2}-{s_{{13}
}}^{2}-4\,{m_{{H}}}^{4}-3\,{m_{{H}}}^{2}s_{{34}}+s_{{13}}s_{{34}}}{
s_{234}s_{24}s_{14}}}\nonumber \\
&&
-4\,{\frac {-{s_{{34}}}^{2}-4\,{m_{{H}}}^
{4}-{s_{{24}}}^{2}+s_{{24}}s_{{34}}+3\,{m_{{H}}}^{2}s_{{24}}-3\,{m_{{H
}}}^{2}s_{{34}}}{s_{134}\,s_{{13}}s_{{23}}}}\nonumber \\
&&
-4\,{\frac {3\,{m_{{H}}
}^{2}s_{{14}}-{s_{{14}}}^{2}+s_{{14}}s_{{34}}
-3\,{m_{{H}}}^{2}s_{{34}}
-4\,{m_{{H}}}^{4}-{s_{{34}}}^{2}}{s_{234}\,s_{{13}}s_{{23}}}}\nonumber \\
&&
-4\,{\frac {s_{{23}}s_{{34}}-4\,{m_{{H}}}^{4}-3\,{m_{{H}}}^{2}s_{{34}}-{s_{
{34}}}^{2}-{s_{{23}}}^{2}+3\,{m_{{H}}}^{2}s_{{23}}}{s_{134}\,s_{{14
}}s_{{24}}}}\nonumber \\
&&
+4\,{\frac {-{s_{{34}}}^{2}-4\,{m_{{H}}}^{4}-{s_{{24}}}^{2
}+s_{{24}}s_{{34}}+3\,{m_{{H}}}^{2}s_{{24}}-3\,{m_{{H}}}^{2}s_{{34}}}{
s_{134}\,s_{{13}}s_{234}}}\nonumber \\
&&
+4\,{\frac {-{s_{{24}}}^{2}-3\,s_{{24}
}s_{{34}}-4\,{m_{{H}}}^{4}-3\,{s_{{34}}}^{2}-6\,{m_{{H}}}^{2}s_{{34}}-
3\,{m_{{H}}}^{2}s_{{24}}}{s_{234}\,s_{134}\,s_{{14}}}}\nonumber \\
&&
+4\,{
\frac {-3\,{s_{{34}}}^{2}-3\,s_{{14}}s_{{34}}-4\,{m_{{H}}}^{4}-3\,{m_{
{H}}}^{2}s_{{14}}-{s_{{14}}}^{2}-6\,{m_{{H}}}^{2}s_{{34}}}{s_{134}
\,s_{234}\,s_{{24}}}}\nonumber \\
&&
+4\,{\frac {3\,{m_{{H}}}^{2}s_{{14}}-{s_{{14}}
}^{2}+s_{{14}}s_{{34}}-3\,{m_{{H}}}^{2}s_{{34}}-4\,{m_{{H}}}^{4}-{s_{{
34}}}^{2}}{s_{234}\,s_{134}\,s_{{23}}}}\nonumber \\
&&
+4\,{\frac {4\,{m_{{H}}}^
{4}+{s_{{23}}}^{2}+3\,{s_{{34}}}^{2}-3\,{m_{{H}}}^{2}s_{{23}}-6\,{m_{{
H}}}^{2}s_{{34}}+3\,s_{{23}}s_{{34}}}{s_{{24}}s_{{13}}s_{{14}}}}\nonumber \\
&&
+4\,{
\frac {4\,{m_{{H}}}^{4}+3\,{s_{{34}}}^{2}+{s_{{24}}}^{2}-6\,{m_{{H}}}^
{2}s_{{34}}+3\,s_{{24}}s_{{34}}-3\,{m_{{H}}}^{2}s_{{24}}}{s_{{13}}s_{{
14}}s_{{23}}}}\nonumber \\
&&	
+4\,{\frac {-3\,{m_{{H}}}^{2}s_{{14}}+4\,{m_{{H}}}^{4}+{
s_{{14}}}^{2}+3\,s_{{14}}s_{{34}}+3\,{s_{{34}}}^{2}-6\,{m_{{H}}}^{2}s_
{{34}}}{s_{{24}}s_{{13}}s_{{23}}}}\nonumber \\
&&	
+4\,{\frac {-6\,{m_{{H}}}^{2}s_{{34}
}+4\,{m_{{H}}}^{4}+3\,s_{{13}}s_{{34}}+{s_{{13}}}^{2}-3\,{m_{{H}}}^{2}
s_{{13}}+3\,{s_{{34}}}^{2}}{s_{{24}}s_{{14}}s_{{23}}}}
\end{eqnarray}

\begin{eqnarray}
 A^{(1)}_{Hb\bar{b}gg} &=&  -16\,{s_{{234}}}^{-1}-32\,{s_{{13}}}^{-1}-24\,{s_{{14}}}^{-1}-16\,{s_{
{134}}}^{-1}-24\,{s_{{24}}}^{-1}-32\,{s_{{23}}}^{-1}\nonumber \\
&&
-16\,{\frac {{m_{{
H}}}^{2}}{{s_{{234}}}^{2}}}-16\,{\frac {{m_{{H}}}^{2}}{{s_{{134}}}^{2}
}}-4\,{\frac {s_{{14}}+2\,s_{{34}}}{s_{{234}}s_{{24}}}}-4\,{\frac {s_{{24}}+2\,s_{{34}}
}{s_{{134}}s_{{14}}}}\nonumber \\
&&
+8\,{\frac {-3
\,s_{{24}}-3\,s_{{34}}+2\,{m_{{H}}}^{2}-3\,s_{{14}}}{s_{{13}}s_{{23}}}
}+4\,{\frac {-5\,s_{{34}}-2\,s_{{13}}+{m_{{H}}}^{2}-2\,s_{{14}}}{s_{{
24}}s_{{23}}}}\nonumber \\
&&
+4\,{\frac {-3\,{m_{{H}}}^{2}-s_{{34}}+s_{{14}}+s_{{24}}
}{s_{{134}}s_{{23}}}}+4\,{\frac {-3\,s_{{24}}-3\,s_{{34}}-3\,{m_{{H}}}
^{2}+4\,s_{{13}}}{s_{{234}}s_{{14}}}}\nonumber \\
&&
-4\,{\frac {-3\,{m_{{H}}}^{2}+8\,
s_{{34}}+6\,s_{{23}}+6\,s_{{14}}}{s_{{24}}s_{{13}}}}-4\,{\frac {3\,{m_
{{H}}}^{2}+3\,s_{{34}}+3\,s_{{14}}-4\,s_{{23}}}{s_{{134}}s_{{24}}}}\nonumber \\
&&
+4
\,{\frac {-3\,{m_{{H}}}^{2}-s_{{34}}+s_{{14}}+s_{{24}}}{s_{{234}}s_{{
13}}}}-4\,{\frac {-3\,{m_{{H}}}^{2}+8\,s_{{34}}+6\,s_{{13}}+6\,s_{{24}
}}{s_{{14}}s_{{23}}}}\nonumber \\
&&
+8\,{\frac {5\,s_{{34}}+4\,{m_{{H}}}^{2}+4\,s_{{
14}}+4\,s_{{24}}}{s_{{134}}s_{{234}}}}-4\,{\frac {-{m_{{H}}}^{2}+s_{{
14}}}{s_{{234}}s_{{23}}}}-4\,{\frac {-{m_{{H}}}^{2}+s_{{24}}}{s_{{134}
}s_{{13}}}}\nonumber \\
&&
+8\,{\frac {2\,{m_{{H}}}^{2}-3\,s_{{23}}-3\,s_{{13}}-3\,s_{
{34}}}{s_{{24}}s_{{14}}}}+4\,{\frac {-2\,s_{{24}}-5\,s_{{34}}-2\,s_{{
23}}+{m_{{H}}}^{2}}{s_{{13}}s_{{14}}}}\nonumber \\
&&
-4\,{\frac {{s_{{13}}}^{2}-{m_{{H}}}^{2}s_{{13}}-
s_{{13}}s_{{34}}+{m_{{H}}}^{2}s_{{34}}+{s_{{34}}}^{2}}{s_{{234}}s_{{24
}}s_{{14}}}}
\nonumber \\
&&
-4\,{\frac {{s_{{24}}}^{2}-{m_{{H}}}^{2}s_{{24}}-s_{{24}}s
_{{34}}+{s_{{34}}}^{2}+{m_{{H}}}^{2}s_{{34}}}{s_{{134}}s_{{13}}s_{{23}
}}}\nonumber \\
&&
-4\,{\frac {-s_{{14}}s_{{34}}+{m_{{H}}}^{2}s_{{34}}+{s_{{34}}}^{2}-
{m_{{H}}}^{2}s_{{14}}+{s_{{14}}}^{2}}{s_{{234}}s_{{13}}s_{{23}}}}
\nonumber \\
&&
-4\,{
\frac {{m_{{H}}}^{2}s_{{34}}+{s_{{23}}}^{2}-s_{{23}}s_{{34}}-{m_{{H}}}
^{2}s_{{23}}+{s_{{34}}}^{2}}{s_{{134}}s_{{14}}s_{{24}}}}\nonumber \\
&&
+4\,{\frac {-s
_{{14}}s_{{34}}+{m_{{H}}}^{2}s_{{34}}+{s_{{34}}}^{2}-{m_{{H}}}^{2}s_{{
14}}+{s_{{14}}}^{2}}{s_{{234}}s_{{134}}s_{{23}}}}
\nonumber \\
&&
+4\,{\frac {{s_{{24}}
}^{2}-{m_{{H}}}^{2}s_{{24}}-s_{{24}}s_{{34}}+{s_{{34}}}^{2}+{m_{{H}}}^
{2}s_{{34}}}{s_{{134}}s_{{13}}s_{{234}}}}\nonumber \\
&&
+4\,{\frac {{s_{{24}}}^{2}+3
\,{s_{{34}}}^{2}+{m_{{H}}}^{2}s_{{24}}+3\,s_{{24}}s_{{34}}+2\,{m_{{H}}
}^{2}s_{{34}}}{s_{{234}}s_{{134}}s_{{14}}}}
\nonumber \\
&&
+4\,{\frac {2\,{m_{{H}}}^{2
}s_{{34}}+{m_{{H}}}^{2}s_{{14}}+{s_{{14}}}^{2}+3\,{s_{{34}}}^{2}+3\,s_
{{14}}s_{{34}}}{s_{{134}}s_{{234}}s_{{24}}}}\nonumber \\
&&
+4\,{\frac {-{s_{{13}}}^{2
}+2\,{m_{{H}}}^{2}s_{{34}}+{m_{{H}}}^{2}s_{{13}}-3\,s_{{13}}s_{{34}}-3
\,{s_{{34}}}^{2}}{s_{{24}}s_{{14}}s_{{23}}}}
\nonumber \\
&&
+4\,{\frac {-3\,s_{{23}}s_
{{34}}+{m_{{H}}}^{2}s_{{23}}-{s_{{23}}}^{2}-3\,{s_{{34}}}^{2}+2\,{m_{{
H}}}^{2}s_{{34}}}{s_{{24}}s_{{13}}s_{{14}}}}\nonumber \\
&&
+4\,{\frac {{m_{{H}}}^{2}s
_{{24}}+2\,{m_{{H}}}^{2}s_{{34}}-3\,s_{{24}}s_{{34}}-{s_{{24}}}^{2}-3
\,{s_{{34}}}^{2}}{s_{{13}}s_{{14}}s_{{23}}}}
\nonumber \\
&&
+4\,{\frac {-3\,{s_{{34}}}
^{2}-{s_{{14}}}^{2}+{m_{{H}}}^{2}s_{{14}}+2\,{m_{{H}}}^{2}s_{{34}}-3\,
s_{{14}}s_{{34}}}{s_{{24}}s_{{13}}s_{{23}}}}
\nonumber \\
&&
+4\,{\frac { \left( -s_{{
34}}+{m_{{H}}}^{2} \right) {s_{{34}}}^{2}}{s_{{13}}s_{{14}}s_{{23}}s_{
{24}}}}
+4\,{\frac { \left( {m_{{H}}}^{2}+s_{{34}} \right) {s_{{34}}}^{
2}}{s_{{234}}s_{{134}}s_{{13}}s_{{23}}}}+4\,{\frac { \left( {m_{{H}}}^
{2}+s_{{34}} \right) {s_{{34}}}^{2}}{s_{{234}}s_{{134}}s_{{14}}s_{{24}
}}}
\end{eqnarray}

\begin{eqnarray}
A^{(2)}_{Hb\bar{b}gg} &=&  -4\,{s_{{23}}}^{-1}-4\,{s_{{13}}}^{-1}-8\,{s_{{134}}}^{-1}-8\,{s_{{234
}}}^{-1}\nonumber \\
&&
+8\,{\frac {{m_{{H}}}^{2}}{{s_{{134}}}^{2}}}+8\,{\frac {{m_{{H
}}}^{2}}{{s_{{234}}}^{2}}}+8\,{\frac {2\,{m_{{H}}}^{2}+2\,s_{{34}}}{s_
{{134}}s_{{234}}}}-4\,{\frac {s_{{34}}+{m_{{H}}}^{2}-s_{{23}}}{s_{{134}}s_{{
14}}}}\nonumber \\
&&
-4\,{\frac {s_{{34}}-s_{{13}}+{m_{{H}}}^{2}}{s_{{234
}}s_{{24}}}}+4\,{\frac {-2\,s_{{34}}+s_{{13}}-s_{{24}}}{s_{{234}}s_{{
14}}}}-4\,{\frac {3\,s_{{34}}+s_{{14}}+s_{{23}}}{s_{{24}}s_{{13}}}}\nonumber \\
&&
+4\,{\frac {s_{{14}}+s_{{24}}-s_{{34}}}{s_{{134}}s_{{23}}}}-4\,{\frac {s
_{{14}}+2\,s_{{34}}-s_{{23}}}{s_{{134}}s_{{24}}}}+4\,{\frac {s_{{14}}+
s_{{24}}-s_{{34}}}{s_{{234}}s_{{13}}}}\nonumber \\
&&
-4\,{\frac {s_{{24}}+3\,s_{{34}}
+s_{{13}}}{s_{{14}}s_{{23}}}}-4\,{\frac {{m_{{H}}}^{2}-s_{{14}}}{s_{{
234}}s_{{23}}}}-4\,{\frac {{m_{{H}}}^{2}-s_{{24}}}{s_{{134}}s_{{13}}}}
\nonumber \\
&&
+4\,{\frac {-s_{{13}}+{m_{{H}}}^{2}-3\,s_{{34}}-s_{{14}}}{s_{{24}}s_{{
23}}}}+4\,{\frac {-s_{{24}}-3\,s_{{34}}+{m_{{H}}}^{2}-s_{{23}}}{s_{{13
}}s_{{14}}}}\nonumber \\
&&
-4\,{\frac {{m_{{H}}}^{2}s_{{34}}-s_{{13}}s_{{34}}+{s_{{34}}}^{2
}}{s_{{234}}s_{{24}}s_{{14}}}}-4\,{\frac {{s_{{34}}}^{2}-s_{{24}}s_{{
34}}+{m_{{H}}}^{2}s_{{34}}}{s_{{134}}s_{{13}}s_{{23}}}}\nonumber \\
&&
-4\,{\frac {{s_
{{34}}}^{2}+{m_{{H}}}^{2}s_{{34}}-s_{{14}}s_{{34}}}{s_{{234}}s_{{13}}s
_{{23}}}}
-4\,{\frac {-s_{{23}}s_{{34}}+{m_{{H}}}^{2}s_{{34}}+{s_{{34}}
}^{2}}{s_{{134}}s_{{14}}s_{{24}}}}\nonumber \\
&&
+4\,{\frac {{s_{{34}}}^{2}-s_{{24}}s
_{{34}}+{m_{{H}}}^{2}s_{{34}}}{s_{{134}}s_{{13}}s_{{234}}}}+4\,{\frac 
{s_{{24}}s_{{34}}+{m_{{H}}}^{2}s_{{34}}+2\,{s_{{34}}}^{2}}{s_{{234}}s_
{{134}}s_{{14}}}}\nonumber \\
&&
+4\,{\frac {s_{{14}}s_{{34}}+{m_{{H}}}^{2}s_{{34}}+2
\,{s_{{34}}}^{2}}{s_{{134}}s_{{234}}s_{{24}}}}+4\,{\frac {{s_{{34}}}^{
2}+{m_{{H}}}^{2}s_{{34}}-s_{{14}}s_{{34}}}{s_{{234}}s_{{134}}s_{{23}}}
}\nonumber \\
&&
+4\,{\frac {-2\,{s_{{34}}}^{2}-s_{{23}}s_{{34}}+{m_{{H}}}^{2}s_{{34}}
}{s_{{24}}s_{{13}}s_{{14}}}}+4\,{\frac {-s_{{24}}s_{{34}}+{m_{{H}}}^{2
}s_{{34}}-2\,{s_{{34}}}^{2}}{s_{{13}}s_{{14}}s_{{23}}}}\nonumber \\
&&
+4\,{\frac {-2
\,{s_{{34}}}^{2}-s_{{14}}s_{{34}}+{m_{{H}}}^{2}s_{{34}}}{s_{{24}}s_{{
13}}s_{{23}}}}+4\,{\frac {-2\,{s_{{34}}}^{2}-s_{{13}}s_{{34}}+{m_{{H}}
}^{2}s_{{34}}}{s_{{24}}s_{{14}}s_{{23}}}}\nonumber \\
&&
+4\,{\frac { \left( -s_{{34}}
+{m_{{H}}}^{2} \right) {s_{{34}}}^{2}}{s_{{13}}s_{{14}}s_{{23}}s_{{24}
}}}+4\,{\frac { \left( {m_{{H}}}^{2}+s_{{34}} \right) {s_{{34}}}^{2}}{
s_{{234}}s_{{134}}s_{{13}}s_{{23}}}}
+4\,{\frac { \left( {m_{{H}}}^{2}+
s_{{34}} \right) {s_{{34}}}^{2}}{s_{{234}}s_{{134}}s_{{14}}s_{{24}}}}
\end{eqnarray}

\begin{eqnarray}
B^{(0)}_{Hb\bar{b}gg} &=& 8\,{s_{{14}}}^{-1}+16\,{s_{{23}}}^{-1}+8\,{s_{{24}}}^{-1}+16\,{s_{{13}
}}^{-1}+40\,{s_{{134}}}^{-1}+40\,{s_{{234}}}^{-1}\nonumber \\
&&
+8\,{\frac {{m_{{H}}}
^{2} \left( -2\,{m_{{H}}}^{2}+s_{{34}}+s_{{14}} \right) }{s_{{234}}s_{
{24}}s_{{13}}}}+8\,{\frac {{m_{{H}}}^{2} \left( -2\,{m_{{H}}}^{2}+s_{{
34}}+s_{{24}} \right) }{s_{{134}}s_{{14}}s_{{23}}}}\nonumber \\
&&
-8\,{\frac {{m_{{H}
}}^{2} \left( 2\,{m_{{H}}}^{2}-s_{{13}}+s_{{24}} \right) }{s_{{234}}s_
{{14}}s_{{23}}}}-8\,{\frac {{m_{{H}}}^{2} \left( 2\,{m_{{H}}}^{2}+s_{{
14}}-s_{{23}} \right) }{s_{{134}}s_{{13}}s_{{24}}}}\nonumber \\
&&
+16\,{\frac {{m_{{H
}}}^{6}}{s_{{134}}s_{{13}}s_{{234}}s_{{24}}}}+16\,{\frac {{m_{{H}}}^{6
}}{s_{{234}}s_{{134}}s_{{14}}s_{{23}}}}
+8\,{\frac {s_{{14}}+2\,{m_{{H}
}}^{2}}{s_{{234}}s_{{13}}}}+8\,{\frac {{m_{{H}}}^{2}}{s_{{134}}s_{{13}
}}}\nonumber \\
&&
-8\,{\frac {-3\,{m_{{H}}}^{2}+s_{{23}}-s_{{24}}-s_{{34}}}{s_{{134}}
s_{{14}}}}+8\,{\frac {2\,s_{{13}}+s_{{34}}+2\,s_{{24}}-{m_{{H}}}^{2}}{
s_{{14}}s_{{23}}}}\nonumber \\
&&
-8\,{\frac {s_{{13}}-s_{{34}}-3\,{m_{{H}}}^{2}-s_{{
14}}}{s_{{234}}s_{{24}}}}+8\,{\frac {{m_{{H}}}^{2}}{s_{{234}}s_{{23}}}
}
-8\,{\frac {{m_{{H}}}^{2}s_{{24}}}{s_{{23}}{s_{{234}}}^{2}}}+8\,{
\frac {{m_{{H}}}^{2}s_{{34}}}{{s_{{134}}}^{2}s_{{14}}}}\nonumber \\
&&
+8\,{\frac {{m_
{{H}}}^{2}s_{{34}}}{{s_{{234}}}^{2}s_{{24}}}}-8\,{\frac {{m_{{H}}}^{2}
s_{{14}}}{{s_{{134}}}^{2}s_{{13}}}}
+8\,{\frac {-{s_{{24}}}^{2}-4\,{m_{
{H}}}^{4}+3\,{m_{{H}}}^{2}s_{{24}}}{s_{{134}}s_{{13}}s_{{234}}}}\nonumber \\
&&
+8\,{
\frac {-2\,s_{{24}}s_{{34}}-3\,{m_{{H}}}^{2}s_{{34}}-4\,{m_{{H}}}^{4}-
{s_{{24}}}^{2}-{s_{{34}}}^{2}-3\,{m_{{H}}}^{2}s_{{24}}}{s_{{234}}s_{{
134}}s_{{14}}}}\nonumber \\
&&
+8\,{\frac {-3\,{m_{{H}}}^{2}s_{{34}}-{s_{{14}}}^{2}-2
\,s_{{14}}s_{{34}}-{s_{{34}}}^{2}-4\,{m_{{H}}}^{4}-3\,{m_{{H}}}^{2}s_{
{14}}}{s_{{134}}s_{{234}}s_{{24}}}}
\nonumber \\
&&
+8\,{\frac {3\,{m_{{H}}}^{2}s_{{14}
}-4\,{m_{{H}}}^{4}-{s_{{14}}}^{2}}{s_{{234}}s_{{134}}s_{{23}}}}
+24\,{
\frac {{m_{{H}}}^{2}}{{s_{{134}}}^{2}}}+24\,{\frac {{m_{{H}}}^{2}}{{s_
{{234}}}^{2}}}\nonumber \\
&&
-8\,{\frac {-s_{{34}}+s_{{13}}-3\,{m_{{H}}}^{2}-2\,s_{{
24}}}{s_{{234}}s_{{14}}}}+8\,{\frac {-{m_{{H}}}^{2}+s_{{34}}+2\,s_{{14
}}+2\,s_{{23}}}{s_{{24}}s_{{13}}}}\nonumber \\
&&
+8\,{\frac {s_{{24}}+2\,{m_{{H}}}^{2
}}{s_{{134}}s_{{23}}}}-8\,{\frac {-s_{{34}}-2\,s_{{14}}-3\,{m_{{H}}}^{
2}+s_{{23}}}{s_{{134}}s_{{24}}}}\nonumber \\
&&
+16\,{\frac {-6\,{m_{{H}}}^{2}-4\,s_{{
24}}-4\,s_{{14}}-3\,s_{{34}}}{s_{{134}}s_{{234}}}}
\end{eqnarray}

\begin{eqnarray}
B^{(1)}_{Hb\bar{b}gg} &=&-8\,{s_{{14}}}^{-1}-16\,{s_{{23}}}^{-1}-8\,{s_{{24}}}^{-1}-16\,{s_{{13}}}^{
-1}-48\,{s_{{134}}}^{-1}-48\,{s_{{234}}}^{-1}\nonumber \\
&&
+8\,{\frac {{m_{{H}}}^{2}
 \left( -2\,s_{{34}}-s_{{14}} \right) }{s_{{234}}s_{{24}}s_{{13}}}}
+8\,{\frac {{m_{{H}}}^{2} \left( -2\,s_{{34}}-s_{{24}} \right) }{s_{{134
}}s_{{14}}s_{{23}}}}-8\,{\frac {{m_{{H}}}^{2} \left( s_{{13}}-2\,s_{{
24}} \right) }{s_{{234}}s_{{14}}s_{{23}}}}\nonumber \\
&&
-8\,{\frac {{m_{{H}}}^{2}
 \left( -2\,s_{{14}}+s_{{23}} \right) }{s_{{134}}s_{{13}}s_{{24}}}}+8
\,{\frac {s_{{13}}}{s_{{234}}s_{{23}}}}-8\,{\frac {-{m_{{H}}}^{2}+s_{{34}}-s_{{23}}}{s_{{134}}s_{{14}}}}\nonumber \\
&&
-8\,{\frac {-s_{{13}}-{m_{{H}}}
^{2}+s_{{34}}}{s_{{234}}s_{{24}}}}+16\,{\frac {{m_{{H}}}^{2}s_{{24}}}{
s_{{23}}{s_{{234}}}^{2}}}-16\,{\frac {{m_{{H}}}^{2}s_{{34}}}{{s_{{134}
}}^{2}s_{{14}}}}\nonumber \\
&&
-16\,{\frac {{m_{{H}}}^{2}s_{{34}}}{{s_{{234}}}^{2}s_{
{24}}}}+16\,{\frac {{m_{{H}}}^{2}s_{{14}}}{{s_{{134}}}^{2}s_{{13}}}}-32\,{\frac {{m_{{H}}}^{2}}{{s_{{134}}}^{2}}}-32\,{\frac {{m_{{H
}}}^{2}}{{s_{{234}}}^{2}}}
\nonumber \\
&&
+8\,{\frac {{s_{{24}}}^{2}-{m_{{H}}}^{2}s_{{24}}}{s_{{134}}s_{{13}}s_{{
234}}}}+8\,{\frac {2\,s_{{24}}s_{{34}}+{s_{{34}}}^{2}+{s_{{24}}}^{2}+{
m_{{H}}}^{2}s_{{24}}+{m_{{H}}}^{2}s_{{34}}}{s_{{234}}s_{{134}}s_{{14}}
}}\nonumber \\
&&
+8\,{\frac {{s_{{34}}}^{2}+2\,s_{{14}}s_{{34}}+{m_{{H}}}^{2}s_{{14}}
+{s_{{14}}}^{2}+{m_{{H}}}^{2}s_{{34}}}{s_{{134}}s_{{234}}s_{{24}}}}+8
\,{\frac {{s_{{14}}}^{2}-{m_{{H}}}^{2}s_{{14}}}{s_{{234}}s_{{134}}s_{{
23}}}}\nonumber \\
&&
-8\,{\frac {-s_{{13}}+{m_{{H}}}^{2}+2\,s_{{
24}}+s_{{34}}}{s_{{234}}s_{{14}}}}+8\,{\frac {-2\,s_{{14}}+{m_{{H}}}^{
2}-s_{{34}}-2\,s_{{23}}}{s_{{24}}s_{{13}}}}\nonumber \\
&&
+8\,{\frac {-2\,s_{{24}}-3
\,{m_{{H}}}^{2}}{s_{{134}}s_{{23}}}}+8\,{\frac {s_{{23}}}{s_{{134}}s_{
{13}}}}-8\,{\frac {{m_{{H}}}^{2}-s_{{23}}+s_{{34}}+2\,s_{{14}}}{s_{{
134}}s_{{24}}}}\nonumber \\
&&
+8\,{\frac {-3\,{m_{{H}}}^{2}-2\,s_{{14}}}{s_{{234}}s_{
{13}}}}+8\,{\frac {-s_{{34}}-2\,s_{{24}}-2\,s_{{13}}+{m_{{H}}}^{2}}{s_
{{14}}s_{{23}}}}\nonumber \\
&&
+16\,{\frac {4\,s_{{14}}+4\,s_{{24}}+2\,{m_{{H}}}^{2}+
3\,s_{{34}}}{s_{{134}}s_{{234}}}} 
\end{eqnarray}

\begin{eqnarray}
B^{(2)}_{Hb\bar{b}gg} &=&
-8\,{\frac {s_{{34}}+s_{{24}}}{s_{{134}}s_{{14}}}}-8\,{s_{{23}}}^{-1}-
8\,{s_{{13}}}^{-1}-16\,{s_{{134}}}^{-1}-16\,{s_{{234}}}^{-1}\nonumber \\
&&
+8\,{
\frac {{m_{{H}}}^{2}s_{{34}}}{s_{{234}}s_{{24}}s_{{13}}}}+8\,{\frac {{
m_{{H}}}^{2}s_{{34}}}{s_{{134}}s_{{14}}s_{{23}}}}-8\,{\frac {{m_{{H}}}
^{2}s_{{24}}}{s_{{234}}s_{{14}}s_{{23}}}}-8\,{\frac {{m_{{H}}}^{2}s_{{
14}}}{s_{{134}}s_{{13}}s_{{24}}}}\nonumber \\
&&
+8\,{\frac {-s_{{13}}+{m_{{H}}}^{2}}{
s_{{234}}s_{{23}}}}-8\,{\frac {s_{{34}}+s_{{14}}}{s_{{234}}s_{{24}}}}-
8\,{\frac {{m_{{H}}}^{2}s_{{24}}}{s_{{23}}{s_{{234}}}^{2}}}\nonumber \\
&&
+8\,{\frac {{m_{{H}}}^{2}s_{{34}}}{{s_{{134}}}^{2}s_{{14}}}}+8\,{\frac {{m_{{H}}}
^{2}s_{{34}}}{{s_{{234}}}^{2}s_{{24}}}}-8\,{\frac {{m_{{H}}}^{2}s_{{14
}}}{{s_{{134}}}^{2}s_{{13}}}}\nonumber \\
&&
+8\,{\frac {-s_{{24}}s_{{34}}+{m_{{H}}}^{
2}s_{{34}}}{s_{{134}}s_{{13}}s_{{234}}}}+8\,{\frac {{s_{{34}}}^{2}+s_{
{24}}s_{{34}}+{m_{{H}}}^{2}s_{{34}}}{s_{{234}}s_{{134}}s_{{14}}}}\nonumber \\
&&
+8\,{
\frac {{s_{{34}}}^{2}+{m_{{H}}}^{2}s_{{34}}+s_{{14}}s_{{34}}}{s_{{134}
}s_{{234}}s_{{24}}}}+8\,{\frac {-s_{{14}}s_{{34}}+{m_{{H}}}^{2}s_{{34}
}}{s_{{234}}s_{{134}}s_{{23}}}}\nonumber \\
&&
+8\,{\frac {{m_{{H}}}^{2}}{{s_{{134}}}^
{2}}}+8\,{\frac {{m_{{H}}}^{2}}{{s_{{234}}}^{2}}}-8\,{\frac {s_{{34}}+
{m_{{H}}}^{2}+s_{{24}}-s_{{13}}}{s_{{234}}s_{{14}}}}\nonumber \\
&&
+8\,{\frac {-s_{{
23}}-s_{{14}}-s_{{34}}+{m_{{H}}}^{2}}{s_{{24}}s_{{13}}}}+8\,{\frac {s_
{{24}}+s_{{14}}}{s_{{134}}s_{{23}}}}+8\,{\frac {-s_{{23}}+{m_{{H}}}^{2
}}{s_{{134}}s_{{13}}}}\nonumber \\
&&
-8\,{\frac {{m_{{H}}}^{2}+s_{{34}}-s_{{23}}+s_{{
14}}}{s_{{134}}s_{{24}}}}+8\,{\frac {s_{{24}}+s_{{14}}}{s_{{234}}s_{{
13}}}}\nonumber \\
&&
+8\,{\frac {{m_{{H}}}^{2}-s_{{34}}-s_{{13}}-s_{{24}}}{s_{{14}}s_
{{23}}}}+16\,{\frac {2\,{m_{{H}}}^{2}+2\,s_{{34}}}{s_{{134}}s_{{234}}}
}
\end{eqnarray}

\begin{eqnarray}
C^{(0)}_{Hb\bar{b}gg} &=&
8\,{\frac {2\,s_{{24}}+4\,{m_{{H}}}^{2}+s_{{14}}-s_{{13}}}{s_{{234}}s_
{{34}}}}-4\,{\frac {-s_{{14}}-2\,s_{{13}}+4\,{m_{{H}}}^{2}-4\,s_{{23}}
}{s_{{24}}s_{{34}}}}\nonumber \\
&&
-4\,{\frac {-2\,s_{{14}}-2\,s_{{23}}+2\,{m_{{H}}}^
{2}-s_{{24}}}{s_{{13}}s_{{34}}}}-4\,{\frac {2\,{m_{{H}}
}^{2}-2\,s_{{13}}-s_{{14}}-2\,s_{{24}}}{s_{{23}}s_{{34}}}}\nonumber \\
&&
+16\,{\frac {-{m_{{H}}}^{2}s_{{24}}-4
\,{m_{{H}}}^{4}-2\,s_{{24}}s_{{14}}-{s_{{14}}}^{2}-{s_{{24}}}^{2}-{m_{
{H}}}^{2}s_{{14}}}{s_{{134}}s_{{34}}s_{{234}}}}\nonumber \\
&&
-4\,{\frac {
-{s_{{24}}}^{2}+2\,{m_{{H}}}^{2}s_{{24}}-2\,{m_{{H}}}^{4}-{s_{{14}}}^{
2}+2\,{m_{{H}}}^{2}s_{{14}}-2\,s_{{24}}s_{{14}}}{s_{{234}}s_{{34}}s_{{
13}}}}\nonumber \\
&&
-4\,{\frac {2\,s_{{24}}s_{{13}}-2\,{m_{{H}}}^{4}-{s_{{24
}}}^{2}-{s_{{13}}}^{2}+2\,{m_{{H}}}^{2}s_{{13}}-2\,{m_{{H}}}^{2}s_{{24
}}}{s_{{234}}s_{{34}}s_{{14}}}}\nonumber \\
&&
-32\,{\frac {s_{{24}}s_{{14}}{m_{{H}}}^{2}}{s_{{134}}{s_{{34}}}^
{2}s_{{234}}}}
-4\,{\frac {-s_{{24}}-2\,s_{{23}}+4\,{m
_{{H}}}^{2}-4\,s_{{13}}}{s_{{14}}s_{{34}}}}\nonumber \\
&&
-4\,{\frac {2\,s_{{14}}s_{{
23}}-{s_{{23}}}^{2}+2\,{m_{{H}}}^{2}s_{{23}}-2\,{m_{{H}}}^{4}-{s_{{14}
}}^{2}-2\,{m_{{H}}}^{2}s_{{14}}}{s_{{134}}s_{{34}}s_{{24}}}}\nonumber \\
&&
-8\,{\frac {-s_{{24}}-4\,{m_{{H}}}^{2}-2\,s_{{14}}+s_{{23}}}{s_{{134}}s_{{
34}}}}+16\,{\frac {{m_{{H}}}^{2}s_{{24}}}{{s_{{234}}}^{2}s_{{34}}}}\nonumber \\
&&
-4\,{\frac {2\,{m_{{H}}}^{2}s_{{24}}+2\,{m_{{H}}}^{2}s_{{13}}-2\,{m_{{H}
}}^{4}-{s_{{24}}}^{2}-2\,s_{{24}}s_{{13}}-{s_{{13}}}^{2}}{s_{{14}}s_{{
23}}s_{{34}}}}\nonumber \\
&&
+16\,{\frac {{s_{{24}}}^{2}{m_{{H}}}^{2}}{{s_{{34}}}^{2}
{s_{{234}}}^{2}}}+16\,{\frac {{m_{{H}}}^{2}s_{{14}}}{{s_{{134}}}^{2}s_
{{34}}}}+16\,{\frac {{s_{{14}}}^{2}{m_{{H}}}^{2}}{{s_{{34}}}^{2}{s_{{
134}}}^{2}}}\nonumber \\
&&
-4\,{\frac {-2\,s_{{14}}s_{{23}}+2\,{m_{{H}}}^{2}s_{{23}}-
2\,{m_{{H}}}^{4}-{s_{{23}}}^{2}+2\,{m_{{H}}}^{2}s_{{14}}-{s_{{14}}}^{2
}}{s_{{24}}s_{{13}}s_{{34}}}}\nonumber \\
&&
-4\,{\frac {-{s_{{24}}}^{2}+2\,{m_{{H}}}^
{2}s_{{24}}-2\,{m_{{H}}}^{4}-{s_{{14}}}^{2}+2\,{m_{{H}}}^{2}s_{{14}}-2
\,s_{{24}}s_{{14}}}{s_{{134}}s_{{34}}s_{{23}}}}
\end{eqnarray}

\begin{eqnarray}
C^{(1)}_{Hb\bar{b}gg} &=& 8\,{\frac {-s_{{14}}-2\,s_{{24}}+s_{{13}}}{s_{{234}}s_{{34}}}}-4\,{
\frac {s_{{24}}+2\,s_{{23}}+s_{{14}}}{s_{{13}}s_{{34}}}}
-4\,{\frac {s_
{{14}}+s_{{24}}+2\,s_{{13}}}{s_{{23}}s_{{34}}}}\nonumber \\
&&+32\,{\frac {s_{{24}}s_
{{14}}{m_{{H}}}^{2}}{s_{{134}}{s_{{34}}}^{2}s_{{234}}}}
-4\,{\frac {2\,
s_{{23}}+4\,s_{{13}}+s_{{24}}}{s_{{14}}s_{{34}}}}-4\,{\frac {-3\,s_{{
14}}s_{{23}}+{s_{{23}}}^{2}+{s_{{14}}}^{2}}{s_{{134}}s_{{34}}s_{{24}}}
}\nonumber \\
&&
-8\,{\frac {2\,s_{{14}}+s_{{24}}-s_{{23}}}{s_{{134}}s_{{34}}}}-16\,{
\frac {{m_{{H}}}^{2}s_{{24}}}{{s_{{234}}}^{2}s_{{34}}}}-4\,{\frac {3\,
s_{{24}}s_{{13}}+{s_{{13}}}^{2}+{s_{{24}}}^{2}}{s_{{14}}s_{{23}}s_{{34
}}}}\nonumber \\
&&
-16\,{\frac {{s_{{24}}}^{2}{m_{{H}}}^{2}}{{s_{{34}}}^{2}{s_{{234}}
}^{2}}}-4\,{\frac {4\,s_{{23}}+s_{{14}}+2\,s_{{13}}}{s_{{24}}s_{{34}}}
}-16\,{\frac {{m_{{H}}}^{2}s_{{14}}}{{s_{{134}}}^{2}s_{{34}}}}\nonumber \\
&&
+16\,{\frac {{m_{{H}}}^{2}s_{{14}}+2\,s_{{24}}s_{{14}}+{m_{{H}}}^{2}s_{{24}}
+{s_{{14}}}^{2}+{s_{{24}}}^{2}}{s_{{134}}s_{{34}}s_{{234}}}}-16\,{\frac {{s_{{14}}}^{2}{m_{{H}}}^{2}}{{s_{{34}}}^{2}{s_{{134}}}^{2}}}
\nonumber \\
&&
-4\,{\frac {{s_{{24}}}^{2}+{s_{{14}}}^{2}+3\,s_{{24}}s_{{14}}}{s_{{234}}
s_{{34}}s_{{13}}}}-4\,{\frac {{s_{{24}}}^{2}+{s_{{14}}}^{2}+3\,s_{{24}}s_{{14}}}{s_{{134}}s_{{34}}s_
{{23}}}}\nonumber \\
&&
-4\,{\frac {{s_{{14}}}^{2}+3\,s_{{14}}s_{{23}}+{s_{{
23}}}^{2}}{s_{{24}}s_{{13}}s_{{34}}}}-4\,{\frac {{s_{{13}}}^{2}+{s_{{
24}}}^{2}-3\,s_{{24}}s_{{13}}}{s_{{234}}s_{{34}}s_{{14}}}}
\end{eqnarray}

\end{appendix}


\bibliographystyle{JHEP}

\end{document}